\documentclass[twocolumn,showpacs,aps,prd,amsmath,amssymb,nofootinbib,nobibnotes,floatfix]{revtex4-1}

\usepackage{graphicx}
\usepackage{float}
\usepackage{bm}
\usepackage[hidelinks]{hyperref}
\usepackage{color}

\usepackage[utf8]{inputenc}
\usepackage[T1]{fontenc}
\usepackage{slashed}

\usepackage{enumerate}

\usepackage{mathalfa,amssymb,amsthm}
\usepackage{amsmath}

\newcommand{\ssec}[1]{\section{#1}}

\newcommand{\beq}{\begin{equation}}
\newcommand{\eeq}{\end{equation}}
\newcommand{\beqa}{\begin{eqnarray}}
\newcommand{\eeqa}{\end{eqnarray}}

\begin{document}

\title{
    Vortex String Formation in Black Hole Superradiance of a Dark Photon 
    with the Higgs Mechanism
   }
\author{William E. East}
\email{weast@perimeterinstitute.ca}
\affiliation{%
Perimeter Institute for Theoretical Physics, Waterloo, Ontario N2L 2Y5, Canada.
}%
\begin{abstract}
Black hole superradiance, which only relies on gravitational interactions, can
provide a powerful probe of the existence of ultralight bosons that are weakly
coupled to ordinary matter. However, as a boson cloud grows through
superradiance, nonlinear effects from interactions with itself or other fields
may become important.  As a representative example of this, we use nonlinear
evolutions to study black hole superradiance of a vector boson that attains a
mass, via a coupling to a complex scalar, through the Higgs mechanism.  For the
cases considered, we find that the superradiant instability can lead to a
transient period where the scalar field reaches its symmetry restoration value,
leading to the formation of closed vortex strings, the temporary disruption of
the exponential growth of the cloud, and an explosive outburst of energy.
After the cloud loses sufficient mass, the superradiant growth resumes, and the
cycle repeats.
Thus, the black hole will be spun down but, potentially, at a much lower rate compared
to when nonlinear effects are unimportant, and with the liberated energy
going primarily into bosonic radiation instead of gravitational waves.
\end{abstract}
\maketitle
\ssec{Introduction}%
A number of extensions to the standard model of particle physics postulate the
existence of ultralight bosons that are weakly coupled to ordinary matter.
This includes the QCD axion~\cite{Peccei:1977hh,Weinberg:1977ma}, the string
axiverse~\cite{Arvanitaki:2009fg,Goodsell:2009xc,Jaeckel:2010ni,Arvanitaki:2010sy},
and dark
photons~\cite{Holdom:1985ag,Pani:2012bp,Graham:2015rva,Agrawal:2018vin}.
For ultralight bosons with Compton wavelengths comparable to the size of
astrophysical black holes, the superradiant instability provides a unique 
observation probe of the existence of these particles. If such a black hole rotates
sufficiently rapidly, it will be unstable to developing a boson
cloud~\cite{Damour:1976,Detweiler:1980uk,Zouros:1979iw}, which may grow to be
up to a few percent of the mass of the black hole, spinning down the black hole in
the process. 

In the absence of other interactions, the saturation of the superradiant
instability comes about through gravitational backreaction.  As the boson cloud
grows, the black hole spins down, and as the rotational frequency of the black
hole approaches that of the bosonic cloud, the instability shuts off, and the
cloud begins to dissipate through gravitational
radiation~\cite{brito_review,East:2017ovw,East:2018glu}. 
This gives rise to a number of potential observational signatures.
One can constrain the existence of ultralight bosons through measurements of
black hole spin inferred from the electromagnetic signatures of accreting
systems or gravitational wave observations of merging binaries~\cite{Arvanitaki:2014wva,Arvanitaki:2016qwi,Baryakhtar:2017ngi,Cardoso:2018tly,Roy:2021uye,Ng:2019jsx,Ng:2020ruv}.
One can search for a gravitational wave signal of 
the oscillating boson cloud, either from resolved or a stochastic background of sources
~\cite{Arvanitaki:2014wva,Arvanitaki:2016qwi,Baryakhtar:2017ngi,allsky_sr_method,allsky_sr_search,CW_galactic,directional_sr_method,Ghosh:2018gaw,Brito:2017wnc,Brito:2017zvb,Tsukada:2018mbp,Tsukada:2020lgt}.
Finally, one can look for the imprint of a boson cloud on the orbital dynamics
of a binary~\cite{Baumann:2019eav,Baumann:2021fkf,Choudhary:2020pxy}. 

However, if the bosonic field has nonlinear interactions with itself or other
matter, these may have important effects on the bosonic cloud before it fully
spins down the black hole, possibly suppressing the spin-down and gravitational
wave observational signatures, but also possibly giving rise to new
observables~\cite{Arvanitaki:2010sy,Yoshino:2012kn,Fukuda:2019ewf,Baryakhtar:2020gao},
including through electromagnetic instabilities and plasma
effects~\cite{Rosa:2017ury,Sen:2018cjt,Boskovic:2018lkj,Caputo:2021efm}.
Compared to the noninteracting case, the role of nonlinear field effects has
been less well studied.  For the case of the axion, one intriguing suggestion
is that, after growing sufficiently large through superradiance, attractive
nonlinear interactions will cause a collapse followed by an energetic outburst,
a phenomenon known as a bosenova~\cite{Arvanitaki:2010sy}.
While numerical simulations~\cite{Yoshino:2012kn,Yoshino:2015nsa} suggest this will happen if
one starts with a sufficiently large boson cloud, perturbative estimates in the
nonrelativistic regime suggest that dissipation through scalar radiation
and/or black hole absorption arising from nonlinear interactions
will halt the growth of the cloud before a bosenova can
occur~\cite{Gruzinov:2016hcq,Baryakhtar:2020gao}
(cf.~Refs.~\cite{Omiya:2020vji,Omiya:2022mwv}).  Part of the challenge in
answering this question is that the timescales associated with the scalar field
superradiant instability rate are prohibitively long for simulations, while
capturing all relevant nonlinear effects with a perturbative analysis is difficult.

In this work, we study black hole superradiance of a vector field that acquires
a mass through the Higgs-mechanism, via a coupling to a complex scalar field.
This is both a physically motivated mechanism for a vector boson to obtain an
ultralight mass~\cite{Goodsell:2009xc,Reece:2018zvv}, and an example of
nonlinear field interactions where it is feasible to perform a full nonlinear
analysis using numerical simulations.  The Abelian Higgs model is also a
prototypical model for cosmic
strings~\cite{1973NuPhB..61...45N,Hindmarsh:1994re,2000csot.book.....V},
analogous to the vortex lines in superconductors, and, recently, has been studied
in the context of dark photon dark
matter~\cite{Long:2019lwl,Redi:2022zkt,Sato:2022jya,upcoming}.

Here, we find that, as the vector boson cloud grows through superradiance, it
drives the scalar field away from its vacuum expectation value (VEV), to
smaller magnitudes in the cloud, eventually leading to the formation of vortex
strings: one dimensional curves where the scalar vanishes. Thus, black hole
superradiance can produce strings, an alternative to cosmic string formation
channels such as phase transitions in the early universe. These string loops
then drive an explosive event, analogous to the bosenova scenario proposed for
the axion, where the cloud is disrupted and loses a significant fraction of its
energy to radiation (as well as absorption by the black hole). After a brief
transient phase, the cloud begins growing again from lower field values, and
the cycle repeats. 

\ssec{Model
   \label{sec:higgs_abelian}
}%
We study an Abelian gauge field in the presence of gravity
that obtains a mass through a Higgs-like coupling to a complex
scalar $\Phi$ with the Lagrangian density
\begin{align}
\mathcal{L}
    =
        \frac{R}{16\pi}
    -\frac{1}{4}F_{ab}F^{ab}
    -\frac{1}{2}D_a\Phi (D^a \Phi)^* -\frac{1}{2}V(|\Phi|^2)
\end{align}
where $F_{ab}:=\nabla_a A_b-\nabla_b A_a$, $D_a:=\nabla_a-igA_a$, 
and we use units with $G=c=\hbar=1$ throughout.
For the potential, we take 
$V(|\Phi|^2)=(\lambda/2)\left(|\Phi|^2-v^2\right)^2$. 
Here 
$g$ and $\lambda$ are coupling constants, and $v$ is the VEV 
of the scalar.  
In this study, we will focus on the regime where nonlinear field
effects become significant before gravitational backreaction is important
and, therefore, will restrict to a fixed black hole spacetime. 

Reviewing the arguments of Ref.~\cite{Fukuda:2019ewf},
we can illustrate some features of this system by writing the complex scalar in terms of a phase $\theta$ and a magnitude 
fluctuation $\rho$ around the VEV, $\Phi=(v+\rho)e^{i \theta}$,
and choosing the unitary gauge where the $U(1)$ symmetry is used to set the
Goldstone boson $\theta=0$.
The vector field equation of motion is then $\nabla_a F^{ab}=\mu^2(1+\rho/v)^2A^b$. Hence, when $\rho\ll v$, the vector field
will act as a Proca field with mass $\mu:=gv$ and can grow exponentially around a black hole through superradiance.
The equation of motion for $\rho$ is $\Box \rho = V'_{\rm eff}(\rho)/2$ with $V_{\rm eff}(\rho)=V(\rho)+\mu^2(1+\rho/v)^2A^2$, where $A^2:=A_aA^a$. 

In this work, we will be interested in the case where $\lambda \gg g^2$, so
that the scalar field is a heavy degree of freedom.  When $A^2\ll A^2_{\rm
c}:=\lambda v^4/\mu^2$, the minimum of the effective potential is at $\rho/v
\approx -A^2/(2A^2_{\rm c})$.  Thus, as the vector field grows through
superradiance, we expect the scalar field to move towards the smaller
magnitude.  When $A^2\geq A^2_{\rm c}$, the minimum of the effective potential
moves to $\Phi=0$, the field value where the spontaneously broken $U(1)$ symmetry is restored, and
the vector becomes effectively massless. However, approaching this point, we
expect strong nonlinear dynamics.

Using a nonrelativistic estimate, we expect the cloud mass at which
$A^2=A^2_{\rm c}$ to be $E/M\sim \alpha^{-4}A^2_{\rm c}$ where $M$ is the
black hole mass and $\alpha:=\mu M$.  Integrating out the scalar will also give
rise to an effective nonlinear term for the vector field 
$\nabla_a F^{ab}\approx \mu^2 (1-A^2/A^2_{\rm c})A^b$, which can lead to vector
radiation which will carry energy away from the system. In the nonrelativistic
and weakly nonlinear regime, we expect the luminosity of this radiation to
scale as $\dot{E}_{\rm rad} \propto \alpha^6 A_{\rm c}^{-4}(E/M)^3$~\cite{Fukuda:2019ewf}. 
This should be compared to the rate at which
energy is extracted from the black hole through the superradiant instability
$\dot{E}_{\rm BH} \propto \alpha^7 (E/M)$~\cite{Baryakhtar:2017ngi}.
Hence, at $A^2=A^2_{\rm c}$, the ratio of the radiation and energy extraction
rate from the black hole should scale as $\dot{E}_{\rm BH}/\dot{E}_{\rm rad}\propto \alpha^9$. 

In the rest of this work, we will use the Lorenz gauge $\nabla_aA^a=0$, since
unitary gauge can be problematic when $|\Phi| \rightarrow 0$.  However, since
in the unitary gauge $\nabla_a([1+\rho/v]^2 A^a)=0$, and we will choose initial
conditions with $\Phi$ real, we expect the two gauges to approximately agree
when $\rho/v\ll 1$ and $A^2\ll A^2_{\rm c}$. 

\ssec{Methodology\label{sec:methodology}}%
We numerically solve the coupled vector-complex scalar equations, 
\begin{equation}
    D_a D^a \Phi = \frac{dV}{d|\Phi|^2}\Phi \ , \nabla^a F_{ab}= -g\times{\rm Im}\left(\Phi^*D_b \Phi\right) \ , 
\label{eqn:eom}
\end{equation}
where, again, $g$ is the gauge coupling constant
in the Lorenz gauge $\nabla_a A^a=0$,
on a fixed black hole spacetime in Kerr-Schild coordinates~\cite{1965cngg.conf..222K}.
The vector field is evolved using the same 3+1 decomposition and constraint damping auxiliary field as in Refs.~\cite{Zilhao:2015tya,East:2017mrj,Helfer:2018qgv}. The complex scalar is evolved as in Ref.~\cite{Siemonsen:2020hcg}.
See appendix for more details on the evolution scheme, numerical resolution, and convergence.

The stress-energy tensor of the system is given by
\beqa
\label{eqn:se}
T_{ab} &=& 
        F_{ac}F_b^c
        -\frac{1}{4}g_{ab}F^{cd}F_{cd} 
        \\ \nonumber
        &&
        +\frac{1}{2}\left [D_a\Phi (D_b \Phi)^*+ \textrm{c.c.}  \right] 
        -\frac{1}{2} g_{ab} \left[ |D_c \Phi|^2  +V(|\Phi|^2) \right]
        . 
\eeqa
We will use the fact that our stationary and axisymmetric spacetime has two
Killing vectors 
$t^a$ and $\phi^a$, 
to define several diagnostic quantities with respect to the stress energy of the system.
The energy and angular momentum are, respectively, given by 
\beq
E := \int -T^t_a t^a N\sqrt{\gamma} d^3x \ , \ J := \int T^t_a \phi^a N\sqrt{\gamma} d^3x  \ , 
\eeq 
where $N$ is the lapse and $\gamma$ is the determinant of the spatial metric.
We will evaluate these quantities outside the black hole horizon, but  
inside some fixed coordinate sphere with size much larger than characteristic boson
cloud size (typically we take $r\geq 50 M$).
Any change in $E$ and $J$ will either be due to a flux through the black hole horizon, or
due to radiation to the wave zone. We will use $\dot{E}_{\rm BH}$ and $\dot{E}_{\rm rad}$
to denote the energy flux calculated at the black hole horizon and in the wave zone, respectively.
We can also divide the energy (and, similarly, the angular momentum) into contributions from
the vector and scalar fields
$E=E_{A}+E_{\Phi}$, corresponding, respectively, to the first
and second lines in Eq.~\ref{eqn:se}. Note that the interaction energy for the two fields is, thus, included in $E_{\Phi}$. 

In this study, we fix the black hole to have mass $M$ and a dimensionless spin
of $a=0.99$. In this case, we can always choose to measure $\Phi$ and $A_a$ in units of
$v$, and the relevant dimensionless parameters are $\alpha:=\mu M$ and $\lambda/g^2=A^2_{\rm c}/v^2$.
Because of the associated computational expense, we will be restricted to considering
cases where $\alpha$ is not too small, and considering large, but not 
extremely large values of $\lambda/g^2$
(though we will comment on how our results extrapolate to other values).  In
particular, we consider $\alpha =0.4$ and $\lambda/g^2=12.5$, 25, and 50. We
also consider $\alpha=0.3$ with $\lambda/g^2=400/9$.  We begin our
evolutions near the end of the weakly nonlinear regime with
$\min(|\Phi|/v)>0.9$.  As described in more detail in the appendix, initial
conditions for a superradiantly growing cloud are constructed by first evolving
an azimuthally symmetric version of the system for a number of $e$ folds, and
using that as the starting point of the full 3D evolution.
We note that, in the absence of nonlinear interactions, the massive vector
field instability growth rate is $\omega_I M=7\times10^{-5}$ and 
$2\times10^{-4}$ for $\alpha=0.3$ and 0.4, respectively~\cite{East:2017mrj}.

\begin{figure}
\begin{center}
\includegraphics[width=\columnwidth,draft=false]{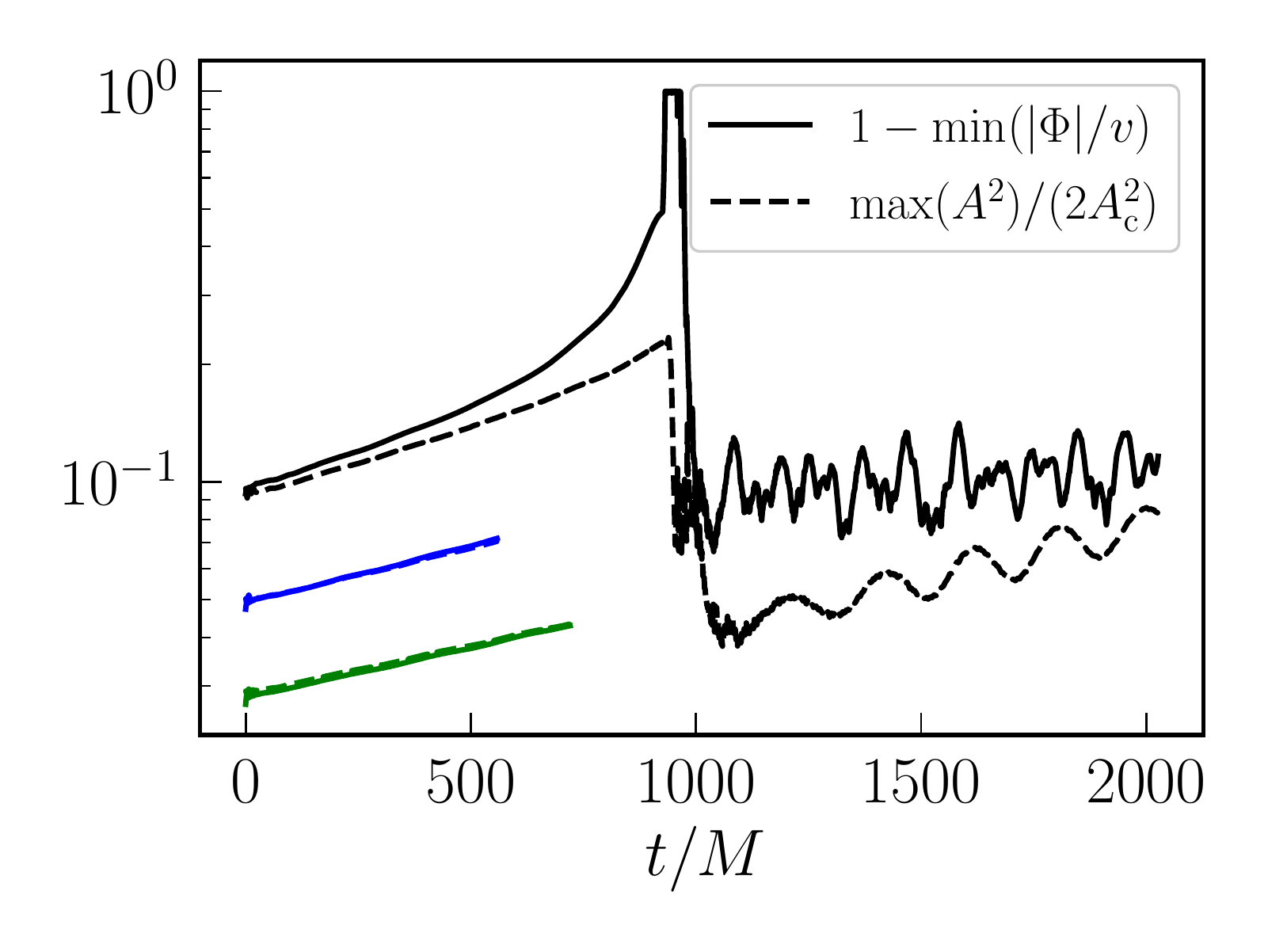}
\end{center}
\caption{
    The displacement of the minimum scalar field value from the VEV (solid
curves) and the maximum vector magnitude (dashed curves) as a function of time
for cases with $\alpha=0.4$ and $\lambda/g^2=25$.  The blue and green curves
have the same parameters as the black curves but have initial values for the
fields that are, respectively, $\approx2$ and $3\times$ smaller.  At lower
field values $\min(|\Phi|/v)\approx 1-\max(A^2)/(2A^2_{\rm c})$. 
\label{fig:field_values}
}
\end{figure}

\ssec{Results}
Our main result is that we find that, after a sufficiently
long period of growth through superradiance, the instability
shuts off with the formation of vortex strings, which eventually
drive the partial disruption of the boson cloud.
In Fig.~\ref{fig:field_values}, we see that, as the vector field grows
exponentially due to superradiance, the minimum value of $|\Phi|$ gets closer to
zero. In particular, as suggested by the simple argument above,
$\min(|\Phi|/v)\sim 1-\max(A^2)/(2A^2_{\rm c})$ in the weakly nonlinear phase.
When $\min(|\Phi|/v)\approx 0.2$, there is a strongly dynamical phase where
$|\Phi|$ quickly approaches zero at certain points in the cloud. 

\begin{figure*}
\begin{center}
\includegraphics[width=0.45\columnwidth,draft=false]{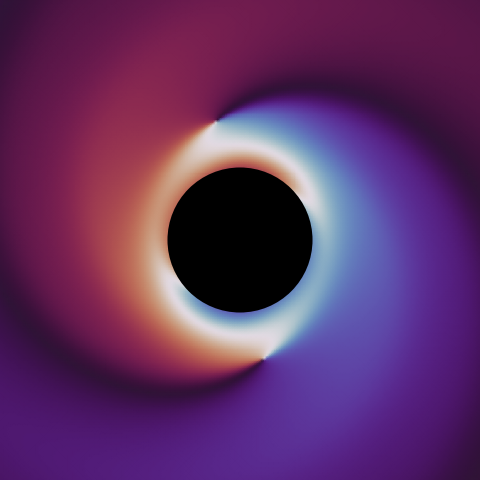}
\includegraphics[width=0.45\columnwidth,draft=false]{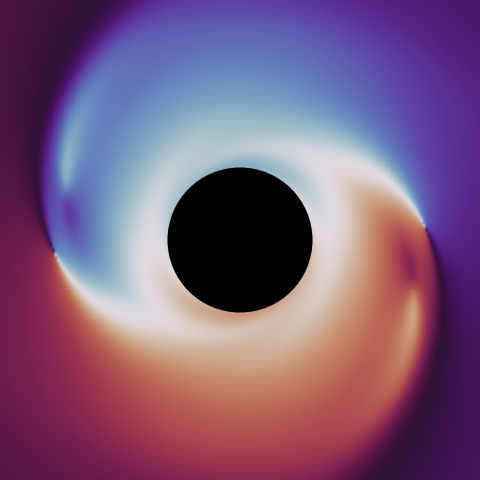}
\includegraphics[width=0.45\columnwidth,draft=false]{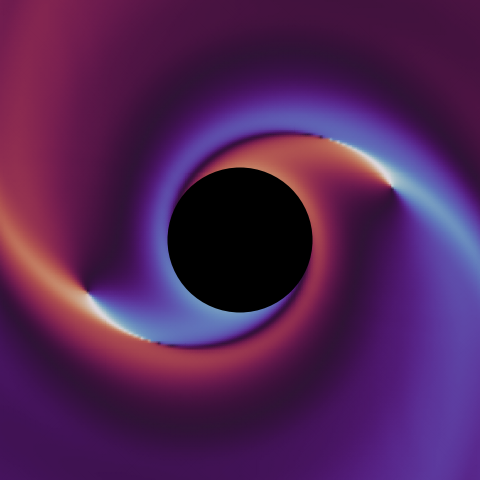}
\includegraphics[width=0.18\columnwidth,draft=false,trim=200 0 0 0, clip]{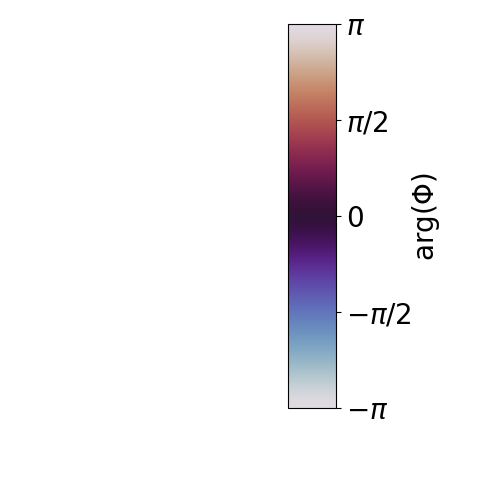}
\includegraphics[width=0.45\columnwidth,draft=false]{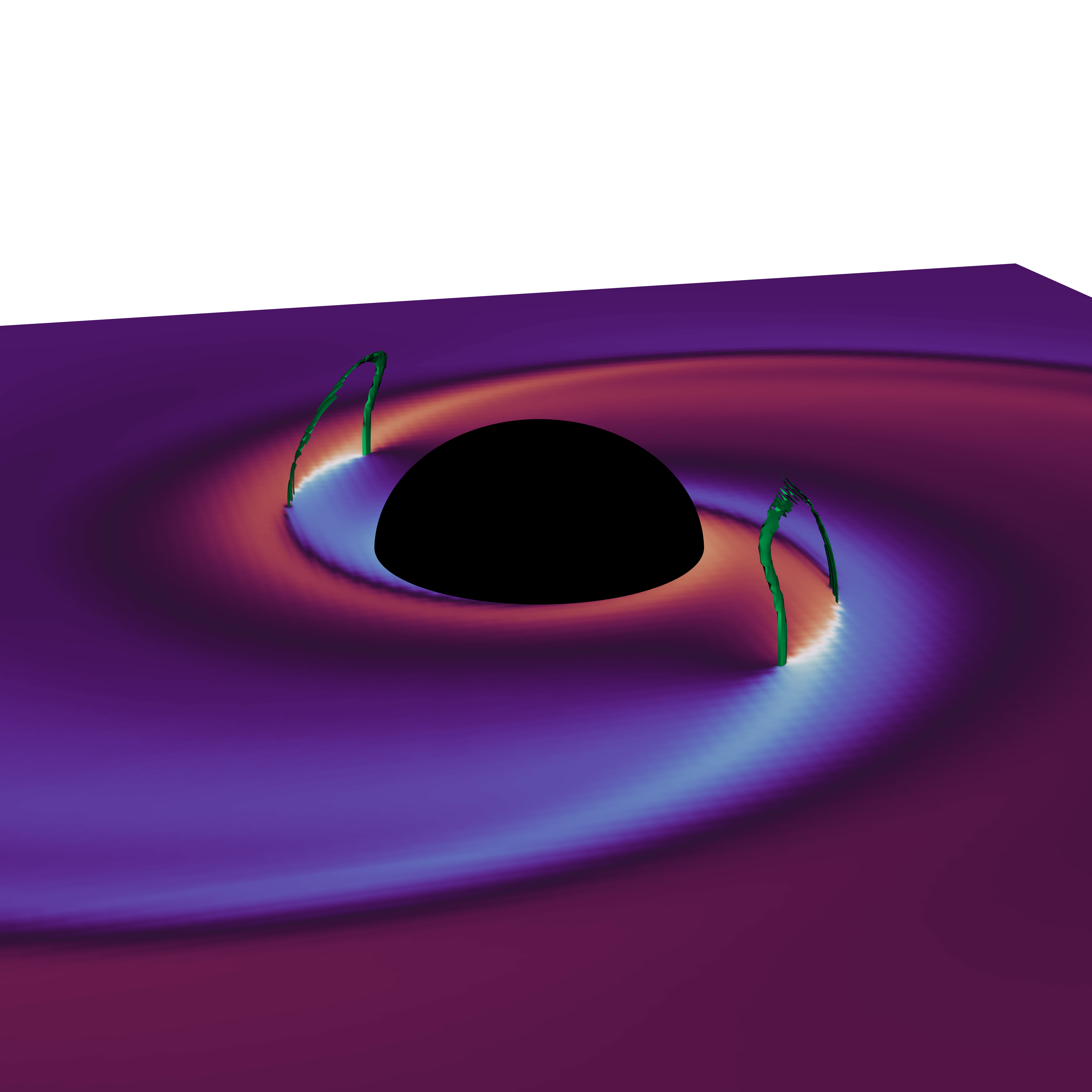}
\includegraphics[width=0.45\columnwidth,draft=false]{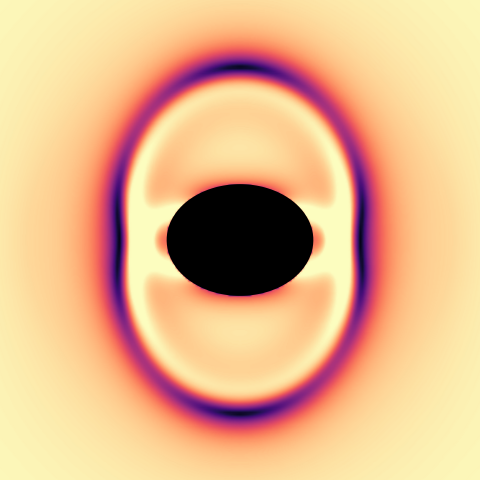}
\includegraphics[width=0.45\columnwidth,draft=false]{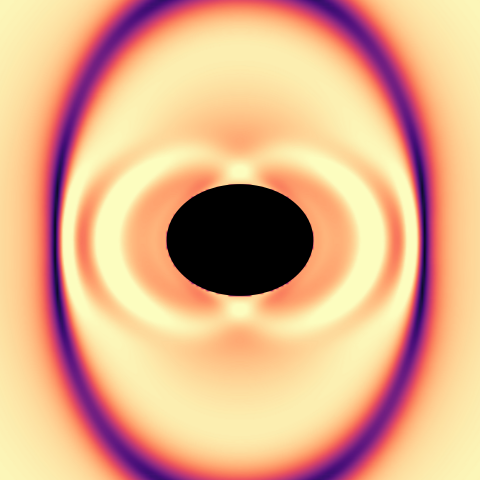}
\includegraphics[width=0.45\columnwidth,draft=false]{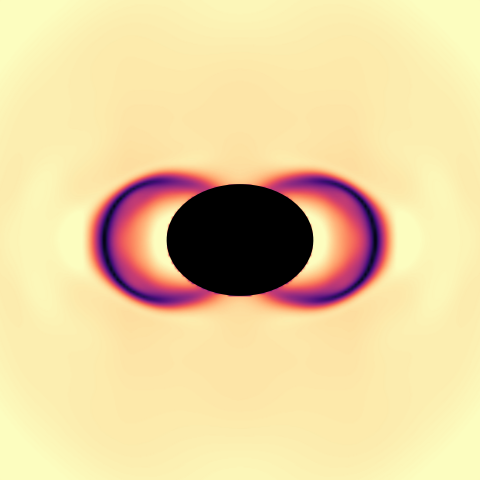}
\includegraphics[width=0.18\columnwidth,draft=false,trim=200 0 0 0, clip]{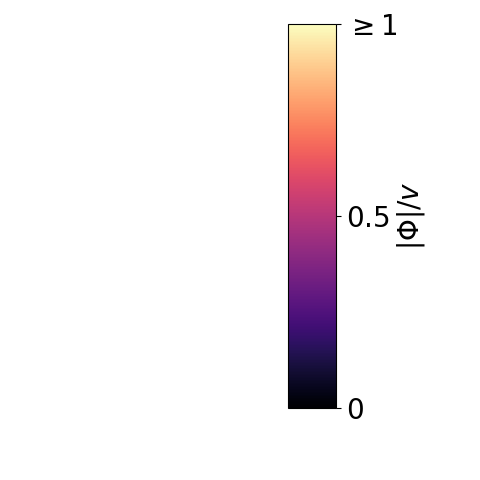}
\includegraphics[width=0.45\columnwidth,draft=false]{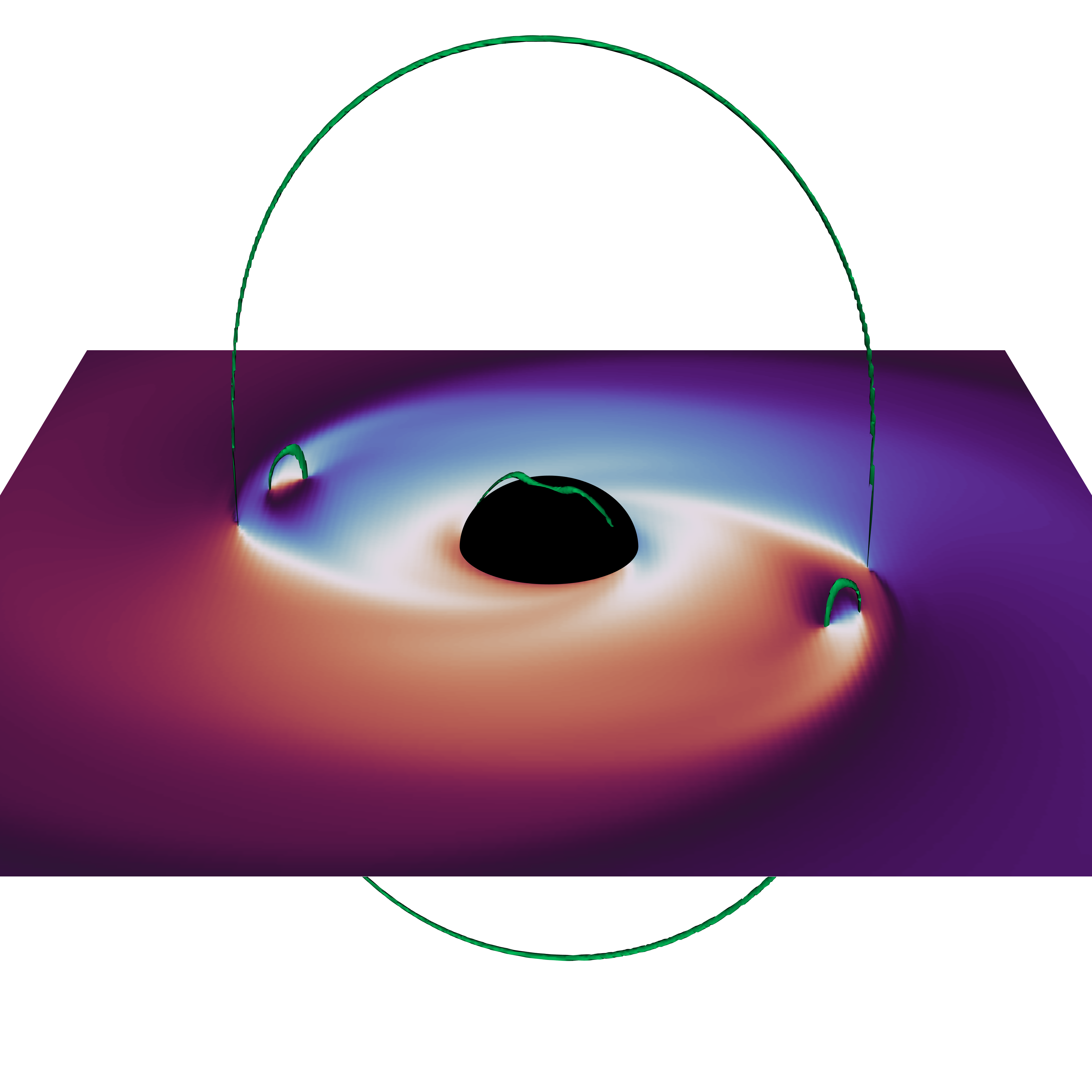}
\end{center}
\caption{
    The three pairs of panels on the left show snapshots of the scalar complex phase $\arg(\Phi)$ in
    the equatorial plane of the black hole (top), and the scalar magnitude $|\Phi|$ in a meridional 
    slice intersecting the black hole and a string at that time, for
    subsequent times (the second and third are, respectively, $7M$ and $24M$
    after the first) following vortex formation in a case with $\alpha=0.4$ and
    $\lambda/g^2=25$. 
    The two rightmost panels show 3D contours of $|\Phi|\approx 0.08v$ in green, indicating the extent of the strings,
    as well as the complex phase in the equatorial plane. 
    The top rightmost panel is the same case and same time as the third column, but shows
    that the nearby pairs of vortices are actually part of the same small loops. (Note that $|\Phi|$ is
    small, but does not go to zero on the black hole horizon here.)
    The bottom rightmost panel is from the case with $\alpha=0.3$.
    For all cases, the scale can be judged by the size of the black hole horizon (black region), 
    which has proper circumference $4\pi M$.
\label{fig:snapshots}
}
\end{figure*}

This phase is
marked by the formation of vortex strings.  Within the cloud, a pair of closed
vortex-antivortex strings form. The electric field of the cloud drives one outward,
while quickly pushing the string with opposite phase winding into the black hole.
As illustrated in the snapshots in Fig.~\ref{fig:snapshots}, for $\alpha=0.4$,
one can see that the
remaining string (first column) roughly spans a meridian outside of the black hole and has
a winding number $|n|=1$ (i.e., the phase of $\Phi$ goes through $2\pi$ when circling the vortex). 
It briefly expands, while continuing to rotate (second column), before tension and gravity 
(coupled with the dissipation of the vector field)
cause it to collapse onto the black
hole as well.  Several short-lived, nonmeridional string loops 
are excited (third column and top rightmost panel), which then fall into the black
hole. Subsequently, as indicated above in Fig.~\ref{fig:field_values},
$|\Phi|/v$ goes above $\sim 0.9$, and there are no vortices.

We also show a snapshot from $\alpha=0.3$ in Fig.~\ref{fig:snapshots}.
This case is similar to the above, except that the main string has a larger spatial
extent, and we also find the formation of additional smaller closed loops
(bottom rightmost panel) which, subsequently, collapse.  

\begin{figure}
\begin{center}
\includegraphics[width=\columnwidth,draft=false]{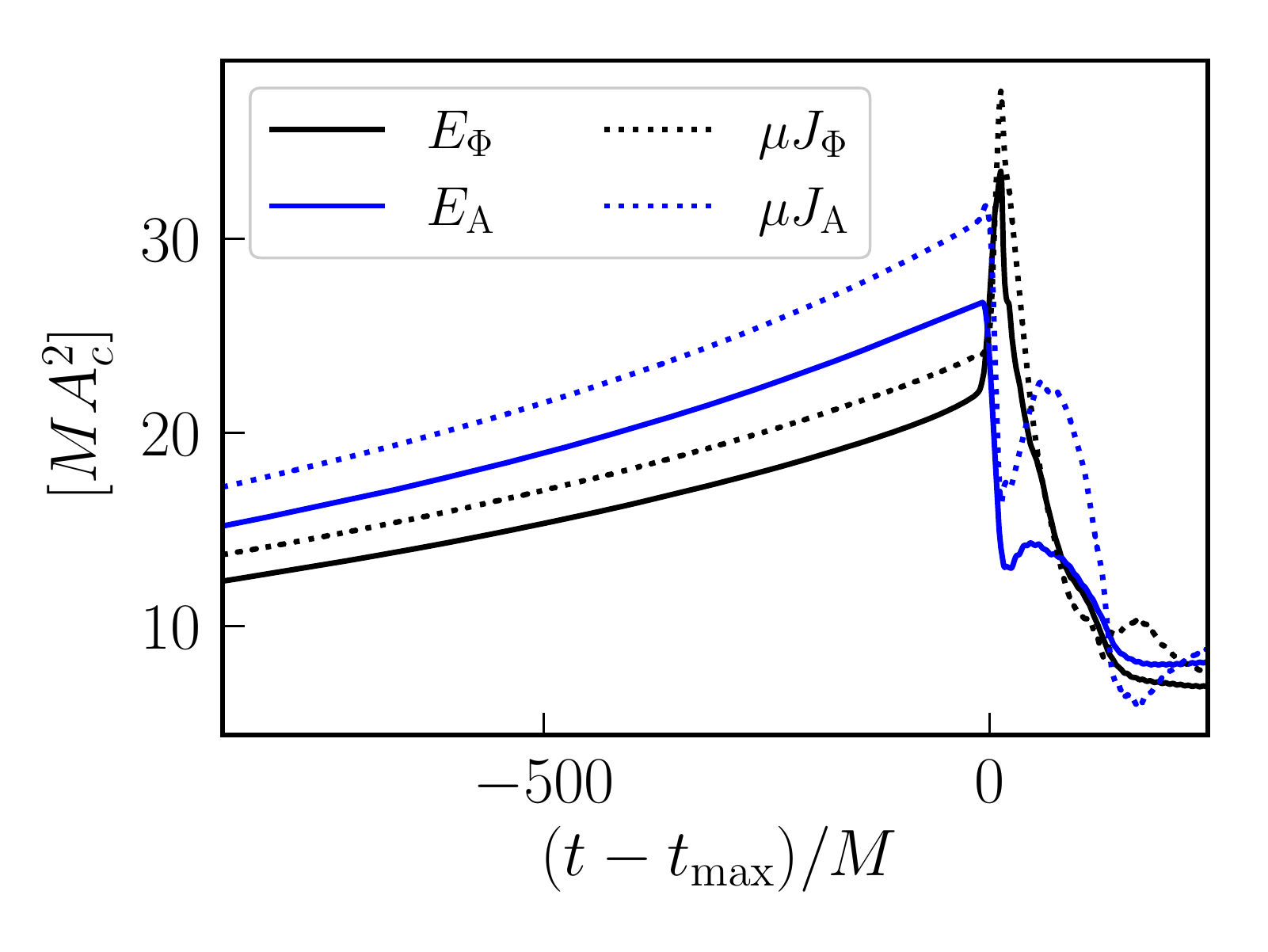}
\includegraphics[width=\columnwidth,draft=false]{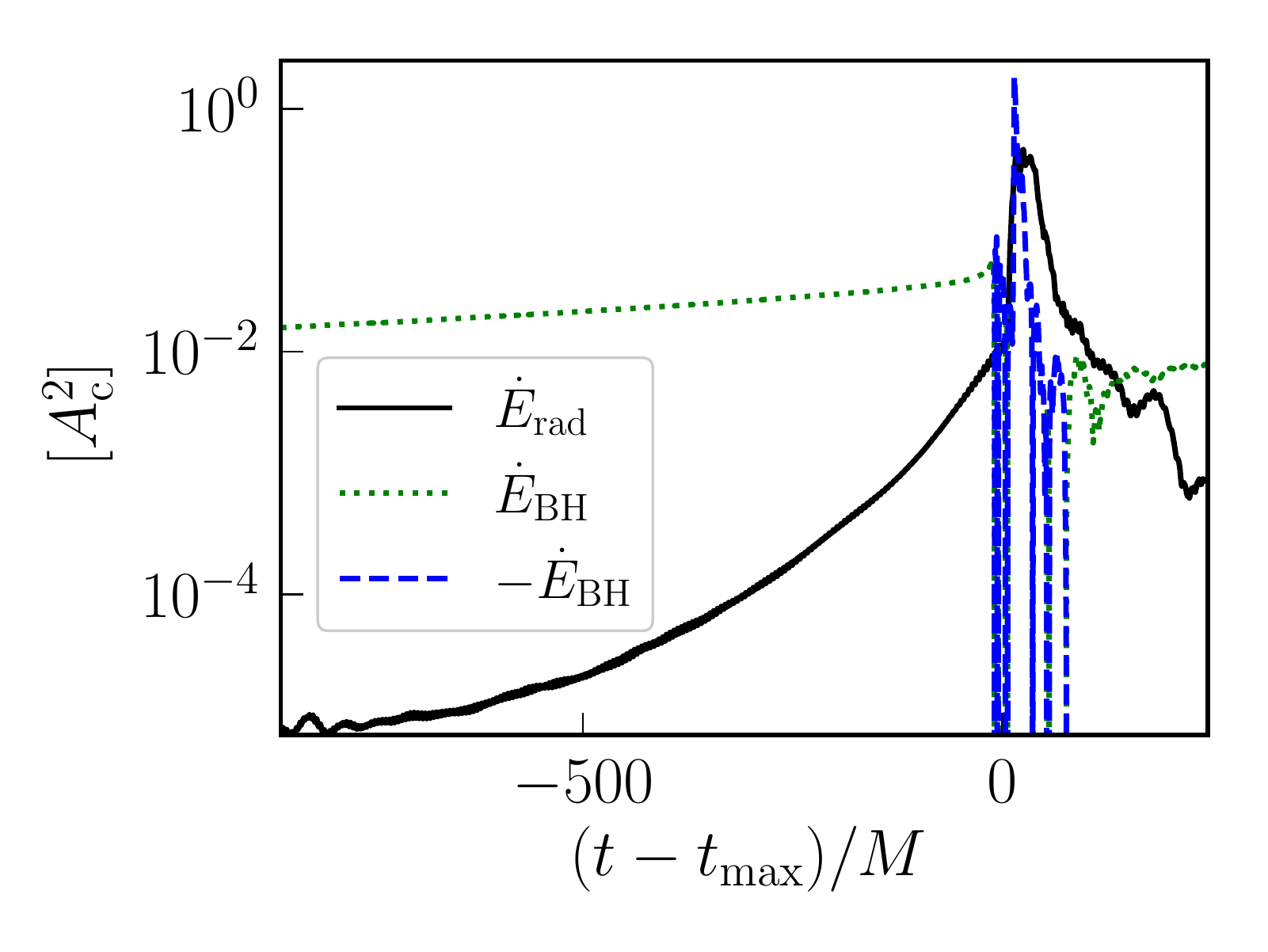}
\end{center}
\caption{
    Top: Energy (solid lines) and angular momentum (dashed lines) as a function of time for $\alpha=0.4$
    and $\lambda/g^2=25$. We show the energy and angular momentum contribution from the scalar field, including the interaction term, 
    $E_{\Phi}$/$J_{\Phi}$, and the vector field (not including terms involving $\Phi$) $E_{A}$/$J_{A}$ separately.
    Bottom: The flux of energy extracted ($\dot{E}_{\rm BH}$) or absorbed ($-\dot{E}_{\rm BH}$) by the black hole
        compared to the radiation luminosity for the same case.
        Here, $t_{\rm max}$ is the time when
            the total energy is maximum.
\label{fig:ej}
}
\end{figure}

In the top panel of Fig.~\ref{fig:ej}, we show the energy and angular momentum
 divided between the vector field contribution and the scalar field
(including the interaction terms with the vector) 
contribution.
As the string vortices form, there is a strong increase in the energy and angular momentum
in the scalar sector, with energy and angular momentum being rapidly drained
out of the vector sector.  
During this phase, the flux of energy out of the black hole switches from 
positive (i.e., superradiance), to strongly negative, as shown in 
the bottom panel of Fig.~\ref{fig:ej}. In addition, in the lead-up to $|\Phi|$ approaching
zero within the cloud, there is a strong rise in the radiation luminosity, much faster
than the $\dot{E}_{\rm rad}\propto E^3$ found in the weakly nonlinear phase,
with a significant burst during the string vortex phase.

\begin{figure}
\begin{center}
\includegraphics[width=\columnwidth,draft=false]{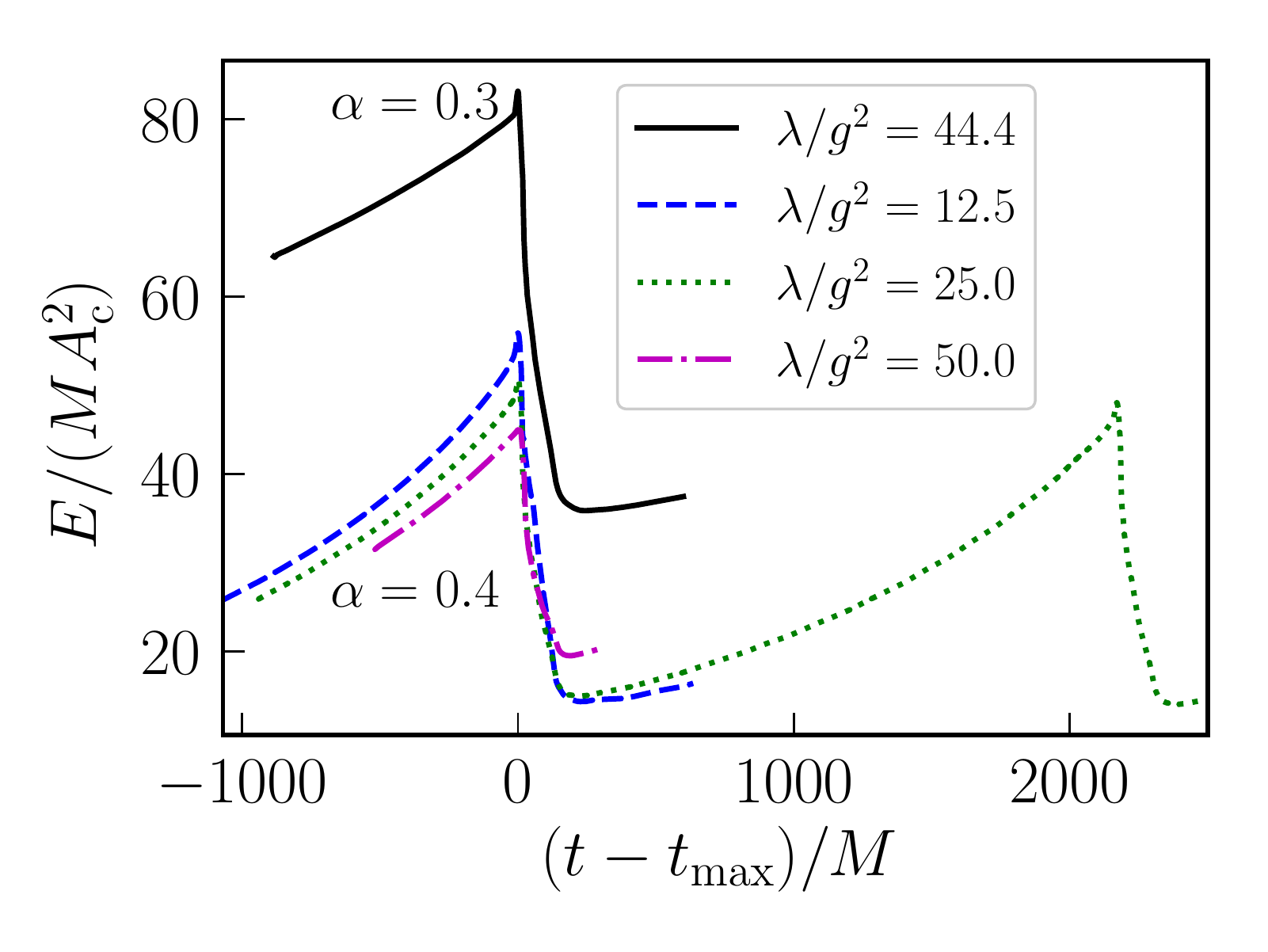}
\end{center}
\caption{
    Total energy as a function of time for cases with $\alpha=0.3$ and $0.4$
    and various values of $\lambda/g^2$. Here $t_{\rm max}$ is the time when 
    the total energy is maximum.
\label{fig:e_lam}
}
\end{figure}

At the end of this strongly dynamical phase, the energy and angular momentum of
the cloud have dropped to roughly $30\%$ of their peak values, as shown in the top
panel of Fig.~\ref{fig:ej}. The majority of this is due to radiation,
with $\sim20\%$ of the energy loss (and even less of the angular momentum loss) 
being due to absorption by the black hole.  Shortly afterwards, the flux of
energy and angular momentum out of the black hole becomes positive again, and the
cloud begins growing exponentially again.  Thus, the cycle will repeat, and 
$\approx2000M$ later, there is another disruption event (see Fig.~\ref{fig:e_lam}).  The ratio of energy to
angular momentum leading up to, and following the cloud disruption, is roughly
the same, and consistent with the linear Proca field frequency $E/J\approx
0.36M^{-1}$.

We also compare several different values of $\lambda$
and $\mu^2$ in Fig.~\ref{fig:e_lam}.  After scaling out the leading order effect, which is that $E$
goes as $A_{\rm c}^2$, we see that there is only mild dependence at fixed
$\alpha$ towards smaller peak values, and higher minimum values following
disruption of the cloud as $A_{\rm c}^2$ is increased.  
As an indication of the numerical error, we estimate the error in the peak value of $E$ 
for $\lambda/g^2=25$ to be $1\%$ (see the appendix for details).
In addition to the
cases with $\alpha=0.4$, in this plot, we also show $\alpha=0.3$, which has
qualitatively similar behavior. As suggested by the nonrelativistic estimates above, the peak energy (scaled by $A_{\rm c}^2$)
is higher, though only $\approx 1.7\times$ higher than $\alpha=0.4$ and fixed $M \lambda v^2/g^2$, which is somewhat less than
the $\alpha^{-4}$ scaling expected in the nonrelativistic limit.
For $\alpha=0.3$, when $\min(|\Phi|)\approx0.2v$, which roughly marks the beginning of the strongly
nonlinear phase, we find that $|\dot{E}_{\rm rad}/\dot{E}_{\rm BH}|\sim 0.01$ (a factor of a few
higher than $\alpha=0.4$). So in this case, the radiation from nonlinear interactions is still
subdominant to the rate at which energy is extracted from the black hole through superradiance. 

\ssec{Discussion and Conclusion}%
We have studied an ultralight boson cloud that grows around a spinning black
hole through superradiance, eventually becoming large enough to strongly
backreact on the Higgs-like scalar field which gave rise to the vector mass.
Restoring physical units for $\alpha=0.4$ assuming $M=60\ M_{\odot}$ (hence,
$\mu=9\times10^{-13}$ eV), this will happen before the cloud reaches saturation
through gravitational interactions (i.e., by spinning down the black hole) when
$v \lambda^{1/4}  \lesssim 10$ MeV or, equivalently, $g \lambda^{-1/4} \gtrsim
10^{-19}$, where $\lambda$ is constrained by unitary considerations to not be too large.  We showed that, in this case, the boson cloud reaches a maximum
energy $E_{\rm max}$ and angular momentum $J_{\rm max}\approx E_{\rm max}/\mu$, with the superradiant growth shutting off as the scalar
field reaches its symmetry restoration value at points within the cloud, giving
rise to string vortices, and with the majority of the energy of the cloud going
into the scalar field. The dynamics of these strings then rapidly dissipate a
significant portion of the cloud, with a fraction $f$ of the angular momentum being
radiated away.
Afterwards, the superradiant growth of the cloud resumes and will persist
until $E_{\rm max}$ and $J_{\rm max}$ are reached again. Thus, angular momentum will continue
to be liberated, but at a slower rate roughly given by $\sim f J_{\rm
max}/[\tau \log(1-f)]$, where $\tau$ is the superradiant energy and angular momentum $e$ folding
time. For the cases considered here (with $\alpha=0.3$--0.4), we found 
$f \gtrsim 0.5$ 
and $E_{\rm max}\sim 50$--$80 M \lambda v^2/g^2 $.  Though we do not consider
gravitational backreaction here, if $\lambda v^2 /g^2$ is sufficiently large so that
$E_{\rm max}$ is non-negligible (but still below the value it would reach in the
Proca limit), we may expect a series of gravitational bursts on timescales of
$\tau$, similar to what was suggested in the bosenova
scenario~\cite{Arvanitaki:2010sy,Yoshino:2012kn,Yoshino:2015nsa}. 
However, we still expect significant bosonic radiation, in contrast to the
Proca limit, where, essentially, all of the rotational energy liberated from the
black hole is emitted as gravitational waves. 

For the cases considered here, we found that the luminosity of the radiation
was subdominant to the energy extraction rate from the black hole due to
superradiance in the lead-up to the strongly nonlinear phase. 
This is consistent with the very recent results of
Ref.~\cite{Clough:2022ygm}, which studied black hole superradiance of a Proca field
with quartic potential
and $\alpha=0.5$, finding that the field growth persists until the evolution equations
break down (such a system could arise by integrating out the scalar in the Abelian Higgs setup considered here).   
 Based on the
relative scalings in the nonrelativistic limit, we expect that,
for sufficiently small values of $\alpha$, the radiation will become dominant
and halt the growth of the cloud before the scalar field is significantly
displaced from its VEV. Though the values of $\alpha$ considered here were too
large to recover the nonrelativistic scaling, we can crudely estimate that this
will happen when $\alpha \lesssim 0.1$.

We have demonstrated a new formation mechanism, distinct from cosmological scenarios,
for forming string loops in the Abelian Higgs model.
In this study, for computational reasons, we have also been restricted to relatively modest values of
$\lambda v^2/ \mu^2 \leq 50$, while in general, given that for astrophysical black
hole superradiance we have $\mu \lesssim 10^{-11}$ eV, it is natural to consider
scenarios where this ratio is many orders of magnitude larger.  In that case,
one may expect the characteristic size of the vortex strings to be much smaller
compared to the boson cloud size and
their density to be larger. An intriguing possibility not covered here is that
this gives rise to a network of interacting strings which undergo many reconnections
and may lead to the ejection of closed strings from the black hole.
This scenario, connections to cosmological scenarios for the dark photon, 
and possible phenomenological implications when considering
a coupling to the standard model are studied in Ref.~\cite{upcoming}.

\ssec{Acknowledgments}
The author thanks Junwu Huang for extensive discussions and
Masha Baryakhtar and David Cyncynates for discussion and comments on a draft
of this work.
The author acknowledges support from an NSERC Discovery grant.
This research was
supported in part by Perimeter Institute for Theoretical Physics. Research at
Perimeter Institute is supported by the Government of Canada through the
Department of Innovation, Science and Economic Development Canada and by the
Province of Ontario through the Ministry of Colleges and Universities.
This research was enabled by computational resources provided by Calcul Québec 
(www.calculquebec.ca) and Compute Canada (www.computecanada.ca), as well
as the Symmetry cluster at Perimeter Institute.

\bibliographystyle{apsrev4-1.bst}
\bibliography{bibliography,ref}

\begin{thebibliography}{62}%
\makeatletter
\providecommand \@ifxundefined [1]{%
 \@ifx{#1\undefined}
}%
\providecommand \@ifnum [1]{%
 \ifnum #1\expandafter \@firstoftwo
 \else \expandafter \@secondoftwo
 \fi
}%
\providecommand \@ifx [1]{%
 \ifx #1\expandafter \@firstoftwo
 \else \expandafter \@secondoftwo
 \fi
}%
\providecommand \natexlab [1]{#1}%
\providecommand \enquote  [1]{``#1''}%
\providecommand \bibnamefont  [1]{#1}%
\providecommand \bibfnamefont [1]{#1}%
\providecommand \citenamefont [1]{#1}%
\providecommand \href@noop [0]{\@secondoftwo}%
\providecommand \href [0]{\begingroup \@sanitize@url \@href}%
\providecommand \@href[1]{\@@startlink{#1}\@@href}%
\providecommand \@@href[1]{\endgroup#1\@@endlink}%
\providecommand \@sanitize@url [0]{\catcode `\\12\catcode `\$12\catcode
  `\&12\catcode `\#12\catcode `\^12\catcode `\_12\catcode `\%12\relax}%
\providecommand \@@startlink[1]{}%
\providecommand \@@endlink[0]{}%
\providecommand \url  [0]{\begingroup\@sanitize@url \@url }%
\providecommand \@url [1]{\endgroup\@href {#1}{\urlprefix }}%
\providecommand \urlprefix  [0]{URL }%
\providecommand \Eprint [0]{\href }%
\providecommand \doibase [0]{http://dx.doi.org/}%
\providecommand \selectlanguage [0]{\@gobble}%
\providecommand \bibinfo  [0]{\@secondoftwo}%
\providecommand \bibfield  [0]{\@secondoftwo}%
\providecommand \translation [1]{[#1]}%
\providecommand \BibitemOpen [0]{}%
\providecommand \bibitemStop [0]{}%
\providecommand \bibitemNoStop [0]{.\EOS\space}%
\providecommand \EOS [0]{\spacefactor3000\relax}%
\providecommand \BibitemShut  [1]{\csname bibitem#1\endcsname}%
\let\auto@bib@innerbib\@empty
\bibitem [{\citenamefont {Peccei}\ and\ \citenamefont
  {Quinn}(1977)}]{Peccei:1977hh}%
  \BibitemOpen
  \bibfield  {author} {\bibinfo {author} {\bibfnamefont {R.}~\bibnamefont
  {Peccei}}\ and\ \bibinfo {author} {\bibfnamefont {H.~R.}\ \bibnamefont
  {Quinn}},\ }\href {\doibase 10.1103/PhysRevLett.38.1440} {\bibfield
  {journal} {\bibinfo  {journal} {Phys. Rev. Lett.}\ }\textbf {\bibinfo
  {volume} {38}},\ \bibinfo {pages} {1440} (\bibinfo {year}
  {1977})}\BibitemShut {NoStop}%
\bibitem [{\citenamefont {Weinberg}(1978)}]{Weinberg:1977ma}%
  \BibitemOpen
  \bibfield  {author} {\bibinfo {author} {\bibfnamefont {S.}~\bibnamefont
  {Weinberg}},\ }\href {\doibase 10.1103/PhysRevLett.40.223} {\bibfield
  {journal} {\bibinfo  {journal} {Phys. Rev. Lett.}\ }\textbf {\bibinfo
  {volume} {40}},\ \bibinfo {pages} {223} (\bibinfo {year} {1978})}\BibitemShut
  {NoStop}%
\bibitem [{\citenamefont {Arvanitaki}\ \emph {et~al.}(2010)\citenamefont
  {Arvanitaki}, \citenamefont {Dimopoulos}, \citenamefont {Dubovsky},
  \citenamefont {Kaloper},\ and\ \citenamefont
  {March-Russell}}]{Arvanitaki:2009fg}%
  \BibitemOpen
  \bibfield  {author} {\bibinfo {author} {\bibfnamefont {A.}~\bibnamefont
  {Arvanitaki}}, \bibinfo {author} {\bibfnamefont {S.}~\bibnamefont
  {Dimopoulos}}, \bibinfo {author} {\bibfnamefont {S.}~\bibnamefont
  {Dubovsky}}, \bibinfo {author} {\bibfnamefont {N.}~\bibnamefont {Kaloper}}, \
  and\ \bibinfo {author} {\bibfnamefont {J.}~\bibnamefont {March-Russell}},\
  }\href {\doibase 10.1103/PhysRevD.81.123530} {\bibfield  {journal} {\bibinfo
  {journal} {Phys. Rev.}\ }\textbf {\bibinfo {volume} {D81}},\ \bibinfo {pages}
  {123530} (\bibinfo {year} {2010})},\ \Eprint {http://arxiv.org/abs/0905.4720}
  {arXiv:0905.4720 [hep-th]} \BibitemShut {NoStop}%
\bibitem [{\citenamefont {Goodsell}\ \emph {et~al.}(2009)\citenamefont
  {Goodsell}, \citenamefont {Jaeckel}, \citenamefont {Redondo},\ and\
  \citenamefont {Ringwald}}]{Goodsell:2009xc}%
  \BibitemOpen
  \bibfield  {author} {\bibinfo {author} {\bibfnamefont {M.}~\bibnamefont
  {Goodsell}}, \bibinfo {author} {\bibfnamefont {J.}~\bibnamefont {Jaeckel}},
  \bibinfo {author} {\bibfnamefont {J.}~\bibnamefont {Redondo}}, \ and\
  \bibinfo {author} {\bibfnamefont {A.}~\bibnamefont {Ringwald}},\ }\href
  {\doibase 10.1088/1126-6708/2009/11/027} {\bibfield  {journal} {\bibinfo
  {journal} {JHEP}\ }\textbf {\bibinfo {volume} {11}},\ \bibinfo {pages} {027}
  (\bibinfo {year} {2009})},\ \Eprint {http://arxiv.org/abs/0909.0515}
  {arXiv:0909.0515 [hep-ph]} \BibitemShut {NoStop}%
\bibitem [{\citenamefont {Jaeckel}\ and\ \citenamefont
  {Ringwald}(2010)}]{Jaeckel:2010ni}%
  \BibitemOpen
  \bibfield  {author} {\bibinfo {author} {\bibfnamefont {J.}~\bibnamefont
  {Jaeckel}}\ and\ \bibinfo {author} {\bibfnamefont {A.}~\bibnamefont
  {Ringwald}},\ }\href {\doibase 10.1146/annurev.nucl.012809.104433} {\bibfield
   {journal} {\bibinfo  {journal} {Ann.Rev.Nucl.Part.Sci.}\ }\textbf {\bibinfo
  {volume} {60}},\ \bibinfo {pages} {405} (\bibinfo {year} {2010})},\ \Eprint
  {http://arxiv.org/abs/1002.0329} {arXiv:1002.0329 [hep-ph]} \BibitemShut
  {NoStop}%
\bibitem [{\citenamefont {Arvanitaki}\ and\ \citenamefont
  {Dubovsky}(2011)}]{Arvanitaki:2010sy}%
  \BibitemOpen
  \bibfield  {author} {\bibinfo {author} {\bibfnamefont {A.}~\bibnamefont
  {Arvanitaki}}\ and\ \bibinfo {author} {\bibfnamefont {S.}~\bibnamefont
  {Dubovsky}},\ }\href {\doibase 10.1103/PhysRevD.83.044026} {\bibfield
  {journal} {\bibinfo  {journal} {Phys. Rev.}\ }\textbf {\bibinfo {volume}
  {D83}},\ \bibinfo {pages} {044026} (\bibinfo {year} {2011})},\ \Eprint
  {http://arxiv.org/abs/1004.3558} {arXiv:1004.3558 [hep-th]} \BibitemShut
  {NoStop}%
\bibitem [{\citenamefont {Holdom}(1986)}]{Holdom:1985ag}%
  \BibitemOpen
  \bibfield  {author} {\bibinfo {author} {\bibfnamefont {B.}~\bibnamefont
  {Holdom}},\ }\href {\doibase 10.1016/0370-2693(86)91377-8} {\bibfield
  {journal} {\bibinfo  {journal} {Phys. Lett.}\ }\textbf {\bibinfo {volume}
  {B166}},\ \bibinfo {pages} {196} (\bibinfo {year} {1986})}\BibitemShut
  {NoStop}%
\bibitem [{\citenamefont {Pani}\ \emph {et~al.}(2012)\citenamefont {Pani},
  \citenamefont {Cardoso}, \citenamefont {Gualtieri}, \citenamefont {Berti},\
  and\ \citenamefont {Ishibashi}}]{Pani:2012bp}%
  \BibitemOpen
  \bibfield  {author} {\bibinfo {author} {\bibfnamefont {P.}~\bibnamefont
  {Pani}}, \bibinfo {author} {\bibfnamefont {V.}~\bibnamefont {Cardoso}},
  \bibinfo {author} {\bibfnamefont {L.}~\bibnamefont {Gualtieri}}, \bibinfo
  {author} {\bibfnamefont {E.}~\bibnamefont {Berti}}, \ and\ \bibinfo {author}
  {\bibfnamefont {A.}~\bibnamefont {Ishibashi}},\ }\href {\doibase
  10.1103/PhysRevD.86.104017} {\bibfield  {journal} {\bibinfo  {journal} {Phys.
  Rev. D}\ }\textbf {\bibinfo {volume} {86}},\ \bibinfo {pages} {104017}
  (\bibinfo {year} {2012})}\BibitemShut {NoStop}%
\bibitem [{\citenamefont {Graham}\ \emph {et~al.}(2016)\citenamefont {Graham},
  \citenamefont {Mardon},\ and\ \citenamefont {Rajendran}}]{Graham:2015rva}%
  \BibitemOpen
  \bibfield  {author} {\bibinfo {author} {\bibfnamefont {P.~W.}\ \bibnamefont
  {Graham}}, \bibinfo {author} {\bibfnamefont {J.}~\bibnamefont {Mardon}}, \
  and\ \bibinfo {author} {\bibfnamefont {S.}~\bibnamefont {Rajendran}},\ }\href
  {\doibase 10.1103/PhysRevD.93.103520} {\bibfield  {journal} {\bibinfo
  {journal} {Phys. Rev. D}\ }\textbf {\bibinfo {volume} {93}},\ \bibinfo
  {pages} {103520} (\bibinfo {year} {2016})},\ \Eprint
  {http://arxiv.org/abs/1504.02102} {arXiv:1504.02102 [hep-ph]} \BibitemShut
  {NoStop}%
\bibitem [{\citenamefont {Agrawal}\ \emph {et~al.}(2020)\citenamefont
  {Agrawal}, \citenamefont {Kitajima}, \citenamefont {Reece}, \citenamefont
  {Sekiguchi},\ and\ \citenamefont {Takahashi}}]{Agrawal:2018vin}%
  \BibitemOpen
  \bibfield  {author} {\bibinfo {author} {\bibfnamefont {P.}~\bibnamefont
  {Agrawal}}, \bibinfo {author} {\bibfnamefont {N.}~\bibnamefont {Kitajima}},
  \bibinfo {author} {\bibfnamefont {M.}~\bibnamefont {Reece}}, \bibinfo
  {author} {\bibfnamefont {T.}~\bibnamefont {Sekiguchi}}, \ and\ \bibinfo
  {author} {\bibfnamefont {F.}~\bibnamefont {Takahashi}},\ }\href {\doibase
  10.1016/j.physletb.2019.135136} {\bibfield  {journal} {\bibinfo  {journal}
  {Phys. Lett. B}\ }\textbf {\bibinfo {volume} {801}},\ \bibinfo {pages}
  {135136} (\bibinfo {year} {2020})},\ \Eprint
  {http://arxiv.org/abs/1810.07188} {arXiv:1810.07188 [hep-ph]} \BibitemShut
  {NoStop}%
\bibitem [{\citenamefont {Damour}\ \emph {et~al.}(1976)\citenamefont {Damour},
  \citenamefont {Deruelle},\ and\ \citenamefont {Ruffini}}]{Damour:1976}%
  \BibitemOpen
  \bibfield  {author} {\bibinfo {author} {\bibfnamefont {T.}~\bibnamefont
  {Damour}}, \bibinfo {author} {\bibfnamefont {N.}~\bibnamefont {Deruelle}}, \
  and\ \bibinfo {author} {\bibfnamefont {R.}~\bibnamefont {Ruffini}},\ }\href
  {\doibase 10.1007/BF02725534} {\bibfield  {journal} {\bibinfo  {journal}
  {Lettere Al Nuovo Cimento Series 2}\ }\textbf {\bibinfo {volume} {15}},\
  \bibinfo {pages} {257} (\bibinfo {year} {1976})}\BibitemShut {NoStop}%
\bibitem [{\citenamefont {Detweiler}(1980)}]{Detweiler:1980uk}%
  \BibitemOpen
  \bibfield  {author} {\bibinfo {author} {\bibfnamefont {S.~L.}\ \bibnamefont
  {Detweiler}},\ }\href {\doibase 10.1103/PhysRevD.22.2323} {\bibfield
  {journal} {\bibinfo  {journal} {Phys.Rev.}\ }\textbf {\bibinfo {volume}
  {D22}},\ \bibinfo {pages} {2323} (\bibinfo {year} {1980})}\BibitemShut
  {NoStop}%
\bibitem [{\citenamefont {Zouros}\ and\ \citenamefont
  {Eardley}(1979)}]{Zouros:1979iw}%
  \BibitemOpen
  \bibfield  {author} {\bibinfo {author} {\bibfnamefont {T.}~\bibnamefont
  {Zouros}}\ and\ \bibinfo {author} {\bibfnamefont {D.}~\bibnamefont
  {Eardley}},\ }\href {\doibase 10.1016/0003-4916(79)90237-9} {\bibfield
  {journal} {\bibinfo  {journal} {Annals Phys.}\ }\textbf {\bibinfo {volume}
  {118}},\ \bibinfo {pages} {139} (\bibinfo {year} {1979})}\BibitemShut
  {NoStop}%
\bibitem [{\citenamefont {Brito}\ \emph {et~al.}(2015)\citenamefont {Brito},
  \citenamefont {Cardoso},\ and\ \citenamefont {Pani}}]{brito_review}%
  \BibitemOpen
  \bibfield  {author} {\bibinfo {author} {\bibfnamefont {R.}~\bibnamefont
  {Brito}}, \bibinfo {author} {\bibfnamefont {V.}~\bibnamefont {Cardoso}}, \
  and\ \bibinfo {author} {\bibfnamefont {P.}~\bibnamefont {Pani}},\ }\href
  {http://stacks.iop.org/0264-9381/32/i=13/a=134001} {\bibfield  {journal}
  {\bibinfo  {journal} {Classical and Quantum Gravity}\ }\textbf {\bibinfo
  {volume} {32}},\ \bibinfo {pages} {134001} (\bibinfo {year}
  {2015})}\BibitemShut {NoStop}%
\bibitem [{\citenamefont {East}\ and\ \citenamefont
  {Pretorius}(2017)}]{East:2017ovw}%
  \BibitemOpen
  \bibfield  {author} {\bibinfo {author} {\bibfnamefont {W.~E.}\ \bibnamefont
  {East}}\ and\ \bibinfo {author} {\bibfnamefont {F.}~\bibnamefont
  {Pretorius}},\ }\href {\doibase 10.1103/PhysRevLett.119.041101} {\bibfield
  {journal} {\bibinfo  {journal} {Phys. Rev. Lett.}\ }\textbf {\bibinfo
  {volume} {119}},\ \bibinfo {pages} {041101} (\bibinfo {year} {2017})},\
  \Eprint {http://arxiv.org/abs/1704.04791} {arXiv:1704.04791 [gr-qc]}
  \BibitemShut {NoStop}%
\bibitem [{\citenamefont {East}(2018)}]{East:2018glu}%
  \BibitemOpen
  \bibfield  {author} {\bibinfo {author} {\bibfnamefont {W.~E.}\ \bibnamefont
  {East}},\ }\href {\doibase 10.1103/PhysRevLett.121.131104} {\bibfield
  {journal} {\bibinfo  {journal} {Phys. Rev. Lett.}\ }\textbf {\bibinfo
  {volume} {121}},\ \bibinfo {pages} {131104} (\bibinfo {year} {2018})},\
  \Eprint {http://arxiv.org/abs/1807.00043} {arXiv:1807.00043 [gr-qc]}
  \BibitemShut {NoStop}%
\bibitem [{\citenamefont {Arvanitaki}\ \emph {et~al.}(2015)\citenamefont
  {Arvanitaki}, \citenamefont {Baryakhtar},\ and\ \citenamefont
  {Huang}}]{Arvanitaki:2014wva}%
  \BibitemOpen
  \bibfield  {author} {\bibinfo {author} {\bibfnamefont {A.}~\bibnamefont
  {Arvanitaki}}, \bibinfo {author} {\bibfnamefont {M.}~\bibnamefont
  {Baryakhtar}}, \ and\ \bibinfo {author} {\bibfnamefont {X.}~\bibnamefont
  {Huang}},\ }\href {\doibase 10.1103/PhysRevD.91.084011} {\bibfield  {journal}
  {\bibinfo  {journal} {Phys. Rev.}\ }\textbf {\bibinfo {volume} {D91}},\
  \bibinfo {pages} {084011} (\bibinfo {year} {2015})},\ \Eprint
  {http://arxiv.org/abs/1411.2263} {arXiv:1411.2263 [hep-ph]} \BibitemShut
  {NoStop}%
\bibitem [{\citenamefont {Arvanitaki}\ \emph {et~al.}(2017)\citenamefont
  {Arvanitaki}, \citenamefont {Baryakhtar}, \citenamefont {Dimopoulos},
  \citenamefont {Dubovsky},\ and\ \citenamefont
  {Lasenby}}]{Arvanitaki:2016qwi}%
  \BibitemOpen
  \bibfield  {author} {\bibinfo {author} {\bibfnamefont {A.}~\bibnamefont
  {Arvanitaki}}, \bibinfo {author} {\bibfnamefont {M.}~\bibnamefont
  {Baryakhtar}}, \bibinfo {author} {\bibfnamefont {S.}~\bibnamefont
  {Dimopoulos}}, \bibinfo {author} {\bibfnamefont {S.}~\bibnamefont
  {Dubovsky}}, \ and\ \bibinfo {author} {\bibfnamefont {R.}~\bibnamefont
  {Lasenby}},\ }\href {\doibase 10.1103/PhysRevD.95.043001} {\bibfield
  {journal} {\bibinfo  {journal} {Phys. Rev.}\ }\textbf {\bibinfo {volume}
  {D95}},\ \bibinfo {pages} {043001} (\bibinfo {year} {2017})},\ \Eprint
  {http://arxiv.org/abs/1604.03958} {arXiv:1604.03958 [hep-ph]} \BibitemShut
  {NoStop}%
\bibitem [{\citenamefont {Baryakhtar}\ \emph {et~al.}(2017)\citenamefont
  {Baryakhtar}, \citenamefont {Lasenby},\ and\ \citenamefont
  {Teo}}]{Baryakhtar:2017ngi}%
  \BibitemOpen
  \bibfield  {author} {\bibinfo {author} {\bibfnamefont {M.}~\bibnamefont
  {Baryakhtar}}, \bibinfo {author} {\bibfnamefont {R.}~\bibnamefont {Lasenby}},
  \ and\ \bibinfo {author} {\bibfnamefont {M.}~\bibnamefont {Teo}},\ }\href
  {\doibase 10.1103/PhysRevD.96.035019} {\bibfield  {journal} {\bibinfo
  {journal} {Phys. Rev. D}\ }\textbf {\bibinfo {volume} {96}},\ \bibinfo
  {pages} {035019} (\bibinfo {year} {2017})},\ \Eprint
  {http://arxiv.org/abs/1704.05081} {arXiv:1704.05081 [hep-ph]} \BibitemShut
  {NoStop}%
\bibitem [{\citenamefont {Cardoso}\ \emph {et~al.}(2018)\citenamefont
  {Cardoso}, \citenamefont {Dias}, \citenamefont {Hartnett}, \citenamefont
  {Middleton}, \citenamefont {Pani},\ and\ \citenamefont
  {Santos}}]{Cardoso:2018tly}%
  \BibitemOpen
  \bibfield  {author} {\bibinfo {author} {\bibfnamefont {V.}~\bibnamefont
  {Cardoso}}, \bibinfo {author} {\bibfnamefont {{\'{O}}.~J.}\ \bibnamefont
  {Dias}}, \bibinfo {author} {\bibfnamefont {G.~S.}\ \bibnamefont {Hartnett}},
  \bibinfo {author} {\bibfnamefont {M.}~\bibnamefont {Middleton}}, \bibinfo
  {author} {\bibfnamefont {P.}~\bibnamefont {Pani}}, \ and\ \bibinfo {author}
  {\bibfnamefont {J.~E.}\ \bibnamefont {Santos}},\ }\href {\doibase
  10.1088/1475-7516/2018/03/043} {\bibfield  {journal} {\bibinfo  {journal}
  {Journal of Cosmology and Astroparticle Physics}\ }\textbf {\bibinfo {volume}
  {2018}},\ \bibinfo {pages} {043} (\bibinfo {year} {2018})}\BibitemShut
  {NoStop}%
\bibitem [{\citenamefont {Roy}\ \emph {et~al.}(2022)\citenamefont {Roy},
  \citenamefont {Vagnozzi},\ and\ \citenamefont {Visinelli}}]{Roy:2021uye}%
  \BibitemOpen
  \bibfield  {author} {\bibinfo {author} {\bibfnamefont {R.}~\bibnamefont
  {Roy}}, \bibinfo {author} {\bibfnamefont {S.}~\bibnamefont {Vagnozzi}}, \
  and\ \bibinfo {author} {\bibfnamefont {L.}~\bibnamefont {Visinelli}},\ }\href
  {\doibase 10.1103/PhysRevD.105.083002} {\bibfield  {journal} {\bibinfo
  {journal} {Phys. Rev. D}\ }\textbf {\bibinfo {volume} {105}},\ \bibinfo
  {pages} {083002} (\bibinfo {year} {2022})},\ \Eprint
  {http://arxiv.org/abs/2112.06932} {arXiv:2112.06932 [astro-ph.HE]}
  \BibitemShut {NoStop}%
\bibitem [{\citenamefont {Ng}\ \emph {et~al.}(2021{\natexlab{a}})\citenamefont
  {Ng}, \citenamefont {Hannuksela}, \citenamefont {Vitale},\ and\ \citenamefont
  {Li}}]{Ng:2019jsx}%
  \BibitemOpen
  \bibfield  {author} {\bibinfo {author} {\bibfnamefont {K.~K.~Y.}\
  \bibnamefont {Ng}}, \bibinfo {author} {\bibfnamefont {O.~A.}\ \bibnamefont
  {Hannuksela}}, \bibinfo {author} {\bibfnamefont {S.}~\bibnamefont {Vitale}},
  \ and\ \bibinfo {author} {\bibfnamefont {T.~G.~F.}\ \bibnamefont {Li}},\
  }\href {\doibase 10.1103/PhysRevD.103.063010} {\bibfield  {journal} {\bibinfo
   {journal} {Phys. Rev. D}\ }\textbf {\bibinfo {volume} {103}},\ \bibinfo
  {pages} {063010} (\bibinfo {year} {2021}{\natexlab{a}})},\ \Eprint
  {http://arxiv.org/abs/1908.02312} {arXiv:1908.02312 [gr-qc]} \BibitemShut
  {NoStop}%
\bibitem [{\citenamefont {Ng}\ \emph {et~al.}(2021{\natexlab{b}})\citenamefont
  {Ng}, \citenamefont {Vitale}, \citenamefont {Hannuksela},\ and\ \citenamefont
  {Li}}]{Ng:2020ruv}%
  \BibitemOpen
  \bibfield  {author} {\bibinfo {author} {\bibfnamefont {K.~K.~Y.}\
  \bibnamefont {Ng}}, \bibinfo {author} {\bibfnamefont {S.}~\bibnamefont
  {Vitale}}, \bibinfo {author} {\bibfnamefont {O.~A.}\ \bibnamefont
  {Hannuksela}}, \ and\ \bibinfo {author} {\bibfnamefont {T.~G.~F.}\
  \bibnamefont {Li}},\ }\href {\doibase 10.1103/PhysRevLett.126.151102}
  {\bibfield  {journal} {\bibinfo  {journal} {Phys. Rev. Lett.}\ }\textbf
  {\bibinfo {volume} {126}},\ \bibinfo {pages} {151102} (\bibinfo {year}
  {2021}{\natexlab{b}})},\ \Eprint {http://arxiv.org/abs/2011.06010}
  {arXiv:2011.06010 [gr-qc]} \BibitemShut {NoStop}%
\bibitem [{\citenamefont {D'Antonio}\ \emph {et~al.}(2018)\citenamefont
  {D'Antonio}, \citenamefont {Palomba}, \citenamefont {Astone}, \citenamefont
  {Frasca}, \citenamefont {Intini}, \citenamefont {La~Rosa}, \citenamefont
  {Leaci}, \citenamefont {Mastrogiovanni}, \citenamefont {Miller},
  \citenamefont {Muciaccia}, \citenamefont {Piccinni},\ and\ \citenamefont
  {Singhal}}]{allsky_sr_method}%
  \BibitemOpen
  \bibfield  {author} {\bibinfo {author} {\bibfnamefont {S.}~\bibnamefont
  {D'Antonio}}, \bibinfo {author} {\bibfnamefont {C.}~\bibnamefont {Palomba}},
  \bibinfo {author} {\bibfnamefont {P.}~\bibnamefont {Astone}}, \bibinfo
  {author} {\bibfnamefont {S.}~\bibnamefont {Frasca}}, \bibinfo {author}
  {\bibfnamefont {G.}~\bibnamefont {Intini}}, \bibinfo {author} {\bibfnamefont
  {I.}~\bibnamefont {La~Rosa}}, \bibinfo {author} {\bibfnamefont
  {P.}~\bibnamefont {Leaci}}, \bibinfo {author} {\bibfnamefont
  {S.}~\bibnamefont {Mastrogiovanni}}, \bibinfo {author} {\bibfnamefont
  {A.}~\bibnamefont {Miller}}, \bibinfo {author} {\bibfnamefont
  {F.}~\bibnamefont {Muciaccia}}, \bibinfo {author} {\bibfnamefont {O.~J.}\
  \bibnamefont {Piccinni}}, \ and\ \bibinfo {author} {\bibfnamefont
  {A.}~\bibnamefont {Singhal}},\ }\href {\doibase 10.1103/PhysRevD.98.103017}
  {\bibfield  {journal} {\bibinfo  {journal} {Phys. Rev. D}\ }\textbf {\bibinfo
  {volume} {98}},\ \bibinfo {pages} {103017} (\bibinfo {year}
  {2018})}\BibitemShut {NoStop}%
\bibitem [{\citenamefont {Palomba}\ \emph {et~al.}(2019)\citenamefont
  {Palomba}, \citenamefont {D'Antonio}, \citenamefont {Astone}, \citenamefont
  {Frasca}, \citenamefont {Intini}, \citenamefont {La~Rosa}, \citenamefont
  {Leaci}, \citenamefont {Mastrogiovanni}, \citenamefont {Miller},
  \citenamefont {Muciaccia}, \citenamefont {Piccinni}, \citenamefont {Rei},\
  and\ \citenamefont {Simula}}]{allsky_sr_search}%
  \BibitemOpen
  \bibfield  {author} {\bibinfo {author} {\bibfnamefont {C.}~\bibnamefont
  {Palomba}}, \bibinfo {author} {\bibfnamefont {S.}~\bibnamefont {D'Antonio}},
  \bibinfo {author} {\bibfnamefont {P.}~\bibnamefont {Astone}}, \bibinfo
  {author} {\bibfnamefont {S.}~\bibnamefont {Frasca}}, \bibinfo {author}
  {\bibfnamefont {G.}~\bibnamefont {Intini}}, \bibinfo {author} {\bibfnamefont
  {I.}~\bibnamefont {La~Rosa}}, \bibinfo {author} {\bibfnamefont
  {P.}~\bibnamefont {Leaci}}, \bibinfo {author} {\bibfnamefont
  {S.}~\bibnamefont {Mastrogiovanni}}, \bibinfo {author} {\bibfnamefont
  {A.~L.}\ \bibnamefont {Miller}}, \bibinfo {author} {\bibfnamefont
  {F.}~\bibnamefont {Muciaccia}}, \bibinfo {author} {\bibfnamefont {O.~J.}\
  \bibnamefont {Piccinni}}, \bibinfo {author} {\bibfnamefont {L.}~\bibnamefont
  {Rei}}, \ and\ \bibinfo {author} {\bibfnamefont {F.}~\bibnamefont {Simula}},\
  }\href {\doibase 10.1103/PhysRevLett.123.171101} {\bibfield  {journal}
  {\bibinfo  {journal} {Phys. Rev. Lett.}\ }\textbf {\bibinfo {volume} {123}},\
  \bibinfo {pages} {171101} (\bibinfo {year} {2019})}\BibitemShut {NoStop}%
\bibitem [{\citenamefont {Zhu}\ \emph {et~al.}(2020)\citenamefont {Zhu},
  \citenamefont {Baryakhtar}, \citenamefont {Papa}, \citenamefont {Tsuna},
  \citenamefont {Kawanaka},\ and\ \citenamefont {Eggenstein}}]{CW_galactic}%
  \BibitemOpen
  \bibfield  {author} {\bibinfo {author} {\bibfnamefont {S.~J.}\ \bibnamefont
  {Zhu}}, \bibinfo {author} {\bibfnamefont {M.}~\bibnamefont {Baryakhtar}},
  \bibinfo {author} {\bibfnamefont {M.~A.}\ \bibnamefont {Papa}}, \bibinfo
  {author} {\bibfnamefont {D.}~\bibnamefont {Tsuna}}, \bibinfo {author}
  {\bibfnamefont {N.}~\bibnamefont {Kawanaka}}, \ and\ \bibinfo {author}
  {\bibfnamefont {H.-B.}\ \bibnamefont {Eggenstein}},\ }\href {\doibase
  10.1103/PhysRevD.102.063020} {\bibfield  {journal} {\bibinfo  {journal}
  {Phys. Rev. D}\ }\textbf {\bibinfo {volume} {102}},\ \bibinfo {pages}
  {063020} (\bibinfo {year} {2020})}\BibitemShut {NoStop}%
\bibitem [{\citenamefont {Isi}\ \emph {et~al.}(2019)\citenamefont {Isi},
  \citenamefont {Sun}, \citenamefont {Brito},\ and\ \citenamefont
  {Melatos}}]{directional_sr_method}%
  \BibitemOpen
  \bibfield  {author} {\bibinfo {author} {\bibfnamefont {M.}~\bibnamefont
  {Isi}}, \bibinfo {author} {\bibfnamefont {L.}~\bibnamefont {Sun}}, \bibinfo
  {author} {\bibfnamefont {R.}~\bibnamefont {Brito}}, \ and\ \bibinfo {author}
  {\bibfnamefont {A.}~\bibnamefont {Melatos}},\ }\href {\doibase
  10.1103/PhysRevD.99.084042} {\bibfield  {journal} {\bibinfo  {journal} {Phys.
  Rev. D}\ }\textbf {\bibinfo {volume} {99}},\ \bibinfo {pages} {084042}
  (\bibinfo {year} {2019})}\BibitemShut {NoStop}%
\bibitem [{\citenamefont {Ghosh}\ \emph {et~al.}(2019)\citenamefont {Ghosh},
  \citenamefont {Berti}, \citenamefont {Brito},\ and\ \citenamefont
  {Richartz}}]{Ghosh:2018gaw}%
  \BibitemOpen
  \bibfield  {author} {\bibinfo {author} {\bibfnamefont {S.}~\bibnamefont
  {Ghosh}}, \bibinfo {author} {\bibfnamefont {E.}~\bibnamefont {Berti}},
  \bibinfo {author} {\bibfnamefont {R.}~\bibnamefont {Brito}}, \ and\ \bibinfo
  {author} {\bibfnamefont {M.}~\bibnamefont {Richartz}},\ }\href {\doibase
  10.1103/PhysRevD.99.104030} {\bibfield  {journal} {\bibinfo  {journal} {Phys.
  Rev. D}\ }\textbf {\bibinfo {volume} {99}},\ \bibinfo {pages} {104030}
  (\bibinfo {year} {2019})},\ \Eprint {http://arxiv.org/abs/1812.01620}
  {arXiv:1812.01620 [gr-qc]} \BibitemShut {NoStop}%
\bibitem [{\citenamefont {Brito}\ \emph
  {et~al.}(2017{\natexlab{a}})\citenamefont {Brito}, \citenamefont {Ghosh},
  \citenamefont {Barausse}, \citenamefont {Berti}, \citenamefont {Cardoso},
  \citenamefont {Dvorkin}, \citenamefont {Klein},\ and\ \citenamefont
  {Pani}}]{Brito:2017wnc}%
  \BibitemOpen
  \bibfield  {author} {\bibinfo {author} {\bibfnamefont {R.}~\bibnamefont
  {Brito}}, \bibinfo {author} {\bibfnamefont {S.}~\bibnamefont {Ghosh}},
  \bibinfo {author} {\bibfnamefont {E.}~\bibnamefont {Barausse}}, \bibinfo
  {author} {\bibfnamefont {E.}~\bibnamefont {Berti}}, \bibinfo {author}
  {\bibfnamefont {V.}~\bibnamefont {Cardoso}}, \bibinfo {author} {\bibfnamefont
  {I.}~\bibnamefont {Dvorkin}}, \bibinfo {author} {\bibfnamefont
  {A.}~\bibnamefont {Klein}}, \ and\ \bibinfo {author} {\bibfnamefont
  {P.}~\bibnamefont {Pani}},\ }\href {\doibase 10.1103/PhysRevLett.119.131101}
  {\bibfield  {journal} {\bibinfo  {journal} {Phys. Rev. Lett.}\ }\textbf
  {\bibinfo {volume} {119}},\ \bibinfo {pages} {131101} (\bibinfo {year}
  {2017}{\natexlab{a}})},\ \Eprint {http://arxiv.org/abs/1706.05097}
  {arXiv:1706.05097 [gr-qc]} \BibitemShut {NoStop}%
\bibitem [{\citenamefont {Brito}\ \emph
  {et~al.}(2017{\natexlab{b}})\citenamefont {Brito}, \citenamefont {Ghosh},
  \citenamefont {Barausse}, \citenamefont {Berti}, \citenamefont {Cardoso},
  \citenamefont {Dvorkin}, \citenamefont {Klein},\ and\ \citenamefont
  {Pani}}]{Brito:2017zvb}%
  \BibitemOpen
  \bibfield  {author} {\bibinfo {author} {\bibfnamefont {R.}~\bibnamefont
  {Brito}}, \bibinfo {author} {\bibfnamefont {S.}~\bibnamefont {Ghosh}},
  \bibinfo {author} {\bibfnamefont {E.}~\bibnamefont {Barausse}}, \bibinfo
  {author} {\bibfnamefont {E.}~\bibnamefont {Berti}}, \bibinfo {author}
  {\bibfnamefont {V.}~\bibnamefont {Cardoso}}, \bibinfo {author} {\bibfnamefont
  {I.}~\bibnamefont {Dvorkin}}, \bibinfo {author} {\bibfnamefont
  {A.}~\bibnamefont {Klein}}, \ and\ \bibinfo {author} {\bibfnamefont
  {P.}~\bibnamefont {Pani}},\ }\href {\doibase 10.1103/PhysRevD.96.064050}
  {\bibfield  {journal} {\bibinfo  {journal} {Phys. Rev. D}\ }\textbf {\bibinfo
  {volume} {96}},\ \bibinfo {pages} {064050} (\bibinfo {year}
  {2017}{\natexlab{b}})},\ \Eprint {http://arxiv.org/abs/1706.06311}
  {arXiv:1706.06311 [gr-qc]} \BibitemShut {NoStop}%
\bibitem [{\citenamefont {Tsukada}\ \emph {et~al.}(2019)\citenamefont
  {Tsukada}, \citenamefont {Callister}, \citenamefont {Matas},\ and\
  \citenamefont {Meyers}}]{Tsukada:2018mbp}%
  \BibitemOpen
  \bibfield  {author} {\bibinfo {author} {\bibfnamefont {L.}~\bibnamefont
  {Tsukada}}, \bibinfo {author} {\bibfnamefont {T.}~\bibnamefont {Callister}},
  \bibinfo {author} {\bibfnamefont {A.}~\bibnamefont {Matas}}, \ and\ \bibinfo
  {author} {\bibfnamefont {P.}~\bibnamefont {Meyers}},\ }\href {\doibase
  10.1103/PhysRevD.99.103015} {\bibfield  {journal} {\bibinfo  {journal} {Phys.
  Rev. D}\ }\textbf {\bibinfo {volume} {99}},\ \bibinfo {pages} {103015}
  (\bibinfo {year} {2019})},\ \Eprint {http://arxiv.org/abs/1812.09622}
  {arXiv:1812.09622 [astro-ph.HE]} \BibitemShut {NoStop}%
\bibitem [{\citenamefont {Tsukada}\ \emph {et~al.}(2021)\citenamefont
  {Tsukada}, \citenamefont {Brito}, \citenamefont {East},\ and\ \citenamefont
  {Siemonsen}}]{Tsukada:2020lgt}%
  \BibitemOpen
  \bibfield  {author} {\bibinfo {author} {\bibfnamefont {L.}~\bibnamefont
  {Tsukada}}, \bibinfo {author} {\bibfnamefont {R.}~\bibnamefont {Brito}},
  \bibinfo {author} {\bibfnamefont {W.~E.}\ \bibnamefont {East}}, \ and\
  \bibinfo {author} {\bibfnamefont {N.}~\bibnamefont {Siemonsen}},\ }\href
  {\doibase 10.1103/PhysRevD.103.083005} {\bibfield  {journal} {\bibinfo
  {journal} {Phys. Rev. D}\ }\textbf {\bibinfo {volume} {103}},\ \bibinfo
  {pages} {083005} (\bibinfo {year} {2021})},\ \Eprint
  {http://arxiv.org/abs/2011.06995} {arXiv:2011.06995 [astro-ph.HE]}
  \BibitemShut {NoStop}%
\bibitem [{\citenamefont {Baumann}\ \emph {et~al.}(2019)\citenamefont
  {Baumann}, \citenamefont {Chia}, \citenamefont {Stout},\ and\ \citenamefont
  {ter Haar}}]{Baumann:2019eav}%
  \BibitemOpen
  \bibfield  {author} {\bibinfo {author} {\bibfnamefont {D.}~\bibnamefont
  {Baumann}}, \bibinfo {author} {\bibfnamefont {H.~S.}\ \bibnamefont {Chia}},
  \bibinfo {author} {\bibfnamefont {J.}~\bibnamefont {Stout}}, \ and\ \bibinfo
  {author} {\bibfnamefont {L.}~\bibnamefont {ter Haar}},\ }\href {\doibase
  10.1088/1475-7516/2019/12/006} {\bibfield  {journal} {\bibinfo  {journal}
  {JCAP}\ }\textbf {\bibinfo {volume} {12}},\ \bibinfo {pages} {006} (\bibinfo
  {year} {2019})},\ \Eprint {http://arxiv.org/abs/1908.10370} {arXiv:1908.10370
  [gr-qc]} \BibitemShut {NoStop}%
\bibitem [{\citenamefont {Baumann}\ \emph {et~al.}(2021)\citenamefont
  {Baumann}, \citenamefont {Bertone}, \citenamefont {Stout},\ and\
  \citenamefont {Tomaselli}}]{Baumann:2021fkf}%
  \BibitemOpen
  \bibfield  {author} {\bibinfo {author} {\bibfnamefont {D.}~\bibnamefont
  {Baumann}}, \bibinfo {author} {\bibfnamefont {G.}~\bibnamefont {Bertone}},
  \bibinfo {author} {\bibfnamefont {J.}~\bibnamefont {Stout}}, \ and\ \bibinfo
  {author} {\bibfnamefont {G.~M.}\ \bibnamefont {Tomaselli}},\ }\href@noop {}
  {\  (\bibinfo {year} {2021})},\ \Eprint {http://arxiv.org/abs/2112.14777}
  {arXiv:2112.14777 [gr-qc]} \BibitemShut {NoStop}%
\bibitem [{\citenamefont {Choudhary}\ \emph {et~al.}(2021)\citenamefont
  {Choudhary}, \citenamefont {Sanchis-Gual}, \citenamefont {Gupta},
  \citenamefont {Degollado}, \citenamefont {Bose},\ and\ \citenamefont
  {Font}}]{Choudhary:2020pxy}%
  \BibitemOpen
  \bibfield  {author} {\bibinfo {author} {\bibfnamefont {S.}~\bibnamefont
  {Choudhary}}, \bibinfo {author} {\bibfnamefont {N.}~\bibnamefont
  {Sanchis-Gual}}, \bibinfo {author} {\bibfnamefont {A.}~\bibnamefont {Gupta}},
  \bibinfo {author} {\bibfnamefont {J.~C.}\ \bibnamefont {Degollado}}, \bibinfo
  {author} {\bibfnamefont {S.}~\bibnamefont {Bose}}, \ and\ \bibinfo {author}
  {\bibfnamefont {J.~A.}\ \bibnamefont {Font}},\ }\href {\doibase
  10.1103/PhysRevD.103.044032} {\bibfield  {journal} {\bibinfo  {journal}
  {Phys. Rev. D}\ }\textbf {\bibinfo {volume} {103}},\ \bibinfo {pages}
  {044032} (\bibinfo {year} {2021})},\ \Eprint
  {http://arxiv.org/abs/2010.00935} {arXiv:2010.00935 [gr-qc]} \BibitemShut
  {NoStop}%
\bibitem [{\citenamefont {Yoshino}\ and\ \citenamefont
  {Kodama}(2012)}]{Yoshino:2012kn}%
  \BibitemOpen
  \bibfield  {author} {\bibinfo {author} {\bibfnamefont {H.}~\bibnamefont
  {Yoshino}}\ and\ \bibinfo {author} {\bibfnamefont {H.}~\bibnamefont
  {Kodama}},\ }\href {\doibase 10.1143/PTP.128.153} {\bibfield  {journal}
  {\bibinfo  {journal} {Prog. Theor. Phys.}\ }\textbf {\bibinfo {volume}
  {128}},\ \bibinfo {pages} {153} (\bibinfo {year} {2012})},\ \Eprint
  {http://arxiv.org/abs/1203.5070} {arXiv:1203.5070 [gr-qc]} \BibitemShut
  {NoStop}%
\bibitem [{\citenamefont {Fukuda}\ and\ \citenamefont
  {Nakayama}(2020)}]{Fukuda:2019ewf}%
  \BibitemOpen
  \bibfield  {author} {\bibinfo {author} {\bibfnamefont {H.}~\bibnamefont
  {Fukuda}}\ and\ \bibinfo {author} {\bibfnamefont {K.}~\bibnamefont
  {Nakayama}},\ }\href {\doibase 10.1007/JHEP01(2020)128} {\bibfield  {journal}
  {\bibinfo  {journal} {JHEP}\ }\textbf {\bibinfo {volume} {01}},\ \bibinfo
  {pages} {128} (\bibinfo {year} {2020})},\ \Eprint
  {http://arxiv.org/abs/1910.06308} {arXiv:1910.06308 [hep-ph]} \BibitemShut
  {NoStop}%
\bibitem [{\citenamefont {Baryakhtar}\ \emph {et~al.}(2021)\citenamefont
  {Baryakhtar}, \citenamefont {Galanis}, \citenamefont {Lasenby},\ and\
  \citenamefont {Simon}}]{Baryakhtar:2020gao}%
  \BibitemOpen
  \bibfield  {author} {\bibinfo {author} {\bibfnamefont {M.}~\bibnamefont
  {Baryakhtar}}, \bibinfo {author} {\bibfnamefont {M.}~\bibnamefont {Galanis}},
  \bibinfo {author} {\bibfnamefont {R.}~\bibnamefont {Lasenby}}, \ and\
  \bibinfo {author} {\bibfnamefont {O.}~\bibnamefont {Simon}},\ }\href
  {\doibase 10.1103/PhysRevD.103.095019} {\bibfield  {journal} {\bibinfo
  {journal} {Phys. Rev. D}\ }\textbf {\bibinfo {volume} {103}},\ \bibinfo
  {pages} {095019} (\bibinfo {year} {2021})},\ \Eprint
  {http://arxiv.org/abs/2011.11646} {arXiv:2011.11646 [hep-ph]} \BibitemShut
  {NoStop}%
\bibitem [{\citenamefont {Rosa}\ and\ \citenamefont
  {Kephart}(2018)}]{Rosa:2017ury}%
  \BibitemOpen
  \bibfield  {author} {\bibinfo {author} {\bibfnamefont {J.~a.~G.}\
  \bibnamefont {Rosa}}\ and\ \bibinfo {author} {\bibfnamefont {T.~W.}\
  \bibnamefont {Kephart}},\ }\href {\doibase 10.1103/PhysRevLett.120.231102}
  {\bibfield  {journal} {\bibinfo  {journal} {Phys. Rev. Lett.}\ }\textbf
  {\bibinfo {volume} {120}},\ \bibinfo {pages} {231102} (\bibinfo {year}
  {2018})},\ \Eprint {http://arxiv.org/abs/1709.06581} {arXiv:1709.06581
  [gr-qc]} \BibitemShut {NoStop}%
\bibitem [{\citenamefont {Sen}(2018)}]{Sen:2018cjt}%
  \BibitemOpen
  \bibfield  {author} {\bibinfo {author} {\bibfnamefont {S.}~\bibnamefont
  {Sen}},\ }\href {\doibase 10.1103/PhysRevD.98.103012} {\bibfield  {journal}
  {\bibinfo  {journal} {Phys. Rev. D}\ }\textbf {\bibinfo {volume} {98}},\
  \bibinfo {pages} {103012} (\bibinfo {year} {2018})},\ \Eprint
  {http://arxiv.org/abs/1805.06471} {arXiv:1805.06471 [hep-ph]} \BibitemShut
  {NoStop}%
\bibitem [{\citenamefont {Boskovic}\ \emph {et~al.}(2019)\citenamefont
  {Boskovic}, \citenamefont {Brito}, \citenamefont {Cardoso}, \citenamefont
  {Ikeda},\ and\ \citenamefont {Witek}}]{Boskovic:2018lkj}%
  \BibitemOpen
  \bibfield  {author} {\bibinfo {author} {\bibfnamefont {M.}~\bibnamefont
  {Boskovic}}, \bibinfo {author} {\bibfnamefont {R.}~\bibnamefont {Brito}},
  \bibinfo {author} {\bibfnamefont {V.}~\bibnamefont {Cardoso}}, \bibinfo
  {author} {\bibfnamefont {T.}~\bibnamefont {Ikeda}}, \ and\ \bibinfo {author}
  {\bibfnamefont {H.}~\bibnamefont {Witek}},\ }\href {\doibase
  10.1103/PhysRevD.99.035006} {\bibfield  {journal} {\bibinfo  {journal} {Phys.
  Rev. D}\ }\textbf {\bibinfo {volume} {99}},\ \bibinfo {pages} {035006}
  (\bibinfo {year} {2019})},\ \Eprint {http://arxiv.org/abs/1811.04945}
  {arXiv:1811.04945 [gr-qc]} \BibitemShut {NoStop}%
\bibitem [{\citenamefont {Caputo}\ \emph {et~al.}(2021)\citenamefont {Caputo},
  \citenamefont {Witte}, \citenamefont {Blas},\ and\ \citenamefont
  {Pani}}]{Caputo:2021efm}%
  \BibitemOpen
  \bibfield  {author} {\bibinfo {author} {\bibfnamefont {A.}~\bibnamefont
  {Caputo}}, \bibinfo {author} {\bibfnamefont {S.~J.}\ \bibnamefont {Witte}},
  \bibinfo {author} {\bibfnamefont {D.}~\bibnamefont {Blas}}, \ and\ \bibinfo
  {author} {\bibfnamefont {P.}~\bibnamefont {Pani}},\ }\href {\doibase
  10.1103/PhysRevD.104.043006} {\bibfield  {journal} {\bibinfo  {journal}
  {Phys. Rev. D}\ }\textbf {\bibinfo {volume} {104}},\ \bibinfo {pages}
  {043006} (\bibinfo {year} {2021})},\ \Eprint
  {http://arxiv.org/abs/2102.11280} {arXiv:2102.11280 [hep-ph]} \BibitemShut
  {NoStop}%
\bibitem [{\citenamefont {Yoshino}\ and\ \citenamefont
  {Kodama}(2015)}]{Yoshino:2015nsa}%
  \BibitemOpen
  \bibfield  {author} {\bibinfo {author} {\bibfnamefont {H.}~\bibnamefont
  {Yoshino}}\ and\ \bibinfo {author} {\bibfnamefont {H.}~\bibnamefont
  {Kodama}},\ }\href {\doibase 10.1088/0264-9381/32/21/214001} {\bibfield
  {journal} {\bibinfo  {journal} {Class. Quant. Grav.}\ }\textbf {\bibinfo
  {volume} {32}},\ \bibinfo {pages} {214001} (\bibinfo {year} {2015})},\
  \Eprint {http://arxiv.org/abs/1505.00714} {arXiv:1505.00714 [gr-qc]}
  \BibitemShut {NoStop}%
\bibitem [{\citenamefont {Gruzinov}(2016)}]{Gruzinov:2016hcq}%
  \BibitemOpen
  \bibfield  {author} {\bibinfo {author} {\bibfnamefont {A.}~\bibnamefont
  {Gruzinov}},\ }\href@noop {} {\  (\bibinfo {year} {2016})},\ \Eprint
  {http://arxiv.org/abs/1604.06422} {arXiv:1604.06422 [astro-ph.HE]}
  \BibitemShut {NoStop}%
\bibitem [{\citenamefont {Omiya}\ \emph {et~al.}(2021)\citenamefont {Omiya},
  \citenamefont {Takahashi},\ and\ \citenamefont {Tanaka}}]{Omiya:2020vji}%
  \BibitemOpen
  \bibfield  {author} {\bibinfo {author} {\bibfnamefont {H.}~\bibnamefont
  {Omiya}}, \bibinfo {author} {\bibfnamefont {T.}~\bibnamefont {Takahashi}}, \
  and\ \bibinfo {author} {\bibfnamefont {T.}~\bibnamefont {Tanaka}},\ }\href
  {\doibase 10.1093/ptep/ptab032} {\bibfield  {journal} {\bibinfo  {journal}
  {PTEP}\ }\textbf {\bibinfo {volume} {2021}},\ \bibinfo {pages} {043E02}
  (\bibinfo {year} {2021})},\ \Eprint {http://arxiv.org/abs/2012.03473}
  {arXiv:2012.03473 [gr-qc]} \BibitemShut {NoStop}%
\bibitem [{\citenamefont {Omiya}\ \emph {et~al.}(2022)\citenamefont {Omiya},
  \citenamefont {Takahashi},\ and\ \citenamefont {Tanaka}}]{Omiya:2022mwv}%
  \BibitemOpen
  \bibfield  {author} {\bibinfo {author} {\bibfnamefont {H.}~\bibnamefont
  {Omiya}}, \bibinfo {author} {\bibfnamefont {T.}~\bibnamefont {Takahashi}}, \
  and\ \bibinfo {author} {\bibfnamefont {T.}~\bibnamefont {Tanaka}},\
  }\href@noop {} {\  (\bibinfo {year} {2022})},\ \Eprint
  {http://arxiv.org/abs/2201.04382} {arXiv:2201.04382 [gr-qc]} \BibitemShut
  {NoStop}%
\bibitem [{\citenamefont {Reece}(2019)}]{Reece:2018zvv}%
  \BibitemOpen
  \bibfield  {author} {\bibinfo {author} {\bibfnamefont {M.}~\bibnamefont
  {Reece}},\ }\href {\doibase 10.1007/JHEP07(2019)181} {\bibfield  {journal}
  {\bibinfo  {journal} {JHEP}\ }\textbf {\bibinfo {volume} {07}},\ \bibinfo
  {pages} {181} (\bibinfo {year} {2019})},\ \Eprint
  {http://arxiv.org/abs/1808.09966} {arXiv:1808.09966 [hep-th]} \BibitemShut
  {NoStop}%
\bibitem [{\citenamefont {{Nielsen}}\ and\ \citenamefont
  {{Olesen}}(1973)}]{1973NuPhB..61...45N}%
  \BibitemOpen
  \bibfield  {author} {\bibinfo {author} {\bibfnamefont {H.~B.}\ \bibnamefont
  {{Nielsen}}}\ and\ \bibinfo {author} {\bibfnamefont {P.}~\bibnamefont
  {{Olesen}}},\ }\href {\doibase 10.1016/0550-3213(73)90350-7} {\bibfield
  {journal} {\bibinfo  {journal} {Nuclear Physics B}\ }\textbf {\bibinfo
  {volume} {61}},\ \bibinfo {pages} {45} (\bibinfo {year} {1973})}\BibitemShut
  {NoStop}%
\bibitem [{\citenamefont {Hindmarsh}\ and\ \citenamefont
  {Kibble}(1995)}]{Hindmarsh:1994re}%
  \BibitemOpen
  \bibfield  {author} {\bibinfo {author} {\bibfnamefont {M.~B.}\ \bibnamefont
  {Hindmarsh}}\ and\ \bibinfo {author} {\bibfnamefont {T.~W.~B.}\ \bibnamefont
  {Kibble}},\ }\href {\doibase 10.1088/0034-4885/58/5/001} {\bibfield
  {journal} {\bibinfo  {journal} {Rept. Prog. Phys.}\ }\textbf {\bibinfo
  {volume} {58}},\ \bibinfo {pages} {477} (\bibinfo {year} {1995})},\ \Eprint
  {http://arxiv.org/abs/hep-ph/9411342} {arXiv:hep-ph/9411342} \BibitemShut
  {NoStop}%
\bibitem [{\citenamefont {{Vilenkin}}\ and\ \citenamefont
  {{Shellard}}(2000)}]{2000csot.book.....V}%
  \BibitemOpen
  \bibfield  {author} {\bibinfo {author} {\bibfnamefont {A.}~\bibnamefont
  {{Vilenkin}}}\ and\ \bibinfo {author} {\bibfnamefont {E.~P.~S.}\ \bibnamefont
  {{Shellard}}},\ }\href@noop {} {\emph {\bibinfo {title} {{Cosmic Strings and
  Other Topological Defects}}}}\ (\bibinfo {year} {2000})\BibitemShut {NoStop}%
\bibitem [{\citenamefont {Long}\ and\ \citenamefont
  {Wang}(2019)}]{Long:2019lwl}%
  \BibitemOpen
  \bibfield  {author} {\bibinfo {author} {\bibfnamefont {A.~J.}\ \bibnamefont
  {Long}}\ and\ \bibinfo {author} {\bibfnamefont {L.-T.}\ \bibnamefont
  {Wang}},\ }\href {\doibase 10.1103/PhysRevD.99.063529} {\bibfield  {journal}
  {\bibinfo  {journal} {Phys. Rev. D}\ }\textbf {\bibinfo {volume} {99}},\
  \bibinfo {pages} {063529} (\bibinfo {year} {2019})},\ \Eprint
  {http://arxiv.org/abs/1901.03312} {arXiv:1901.03312 [hep-ph]} \BibitemShut
  {NoStop}%
\bibitem [{\citenamefont {Redi}\ and\ \citenamefont
  {Tesi}(2022)}]{Redi:2022zkt}%
  \BibitemOpen
  \bibfield  {author} {\bibinfo {author} {\bibfnamefont {M.}~\bibnamefont
  {Redi}}\ and\ \bibinfo {author} {\bibfnamefont {A.}~\bibnamefont {Tesi}},\
  }\href@noop {} {\  (\bibinfo {year} {2022})},\ \Eprint
  {http://arxiv.org/abs/2204.14274} {arXiv:2204.14274 [hep-ph]} \BibitemShut
  {NoStop}%
\bibitem [{\citenamefont {Sato}\ \emph {et~al.}(2022)\citenamefont {Sato},
  \citenamefont {Takahashi},\ and\ \citenamefont {Yamada}}]{Sato:2022jya}%
  \BibitemOpen
  \bibfield  {author} {\bibinfo {author} {\bibfnamefont {T.}~\bibnamefont
  {Sato}}, \bibinfo {author} {\bibfnamefont {F.}~\bibnamefont {Takahashi}}, \
  and\ \bibinfo {author} {\bibfnamefont {M.}~\bibnamefont {Yamada}},\
  }\href@noop {} {\  (\bibinfo {year} {2022})},\ \Eprint
  {http://arxiv.org/abs/2204.11896} {arXiv:2204.11896 [hep-ph]} \BibitemShut
  {NoStop}%
\bibitem [{\citenamefont {East}\ and\ \citenamefont {Huang}(2022)}]{upcoming}%
  \BibitemOpen
  \bibfield  {author} {\bibinfo {author} {\bibfnamefont {W.~E.}\ \bibnamefont
  {East}}\ and\ \bibinfo {author} {\bibfnamefont {J.}~\bibnamefont {Huang}},\
  }\href@noop {} {\  (\bibinfo {year} {2022})},\ \Eprint
  {http://arxiv.org/abs/2206.12432} {arXiv:2206.12432 [hep-ph]} \BibitemShut
  {NoStop}%
\bibitem [{\citenamefont {{Kerr}}\ and\ \citenamefont
  {{Schild}}(1965)}]{1965cngg.conf..222K}%
  \BibitemOpen
  \bibfield  {author} {\bibinfo {author} {\bibfnamefont {R.~P.}\ \bibnamefont
  {{Kerr}}}\ and\ \bibinfo {author} {\bibfnamefont {A.}~\bibnamefont
  {{Schild}}},\ }in\ \href@noop {} {\emph {\bibinfo {booktitle} {IV Centenario
  Della Nascita di Galileo Galilei}}}\ (\bibinfo {year} {1965})\ p.\ \bibinfo
  {pages} {222}\BibitemShut {NoStop}%
\bibitem [{\citenamefont {Zilhão}\ \emph {et~al.}(2015)\citenamefont
  {Zilhão}, \citenamefont {Witek},\ and\ \citenamefont
  {Cardoso}}]{Zilhao:2015tya}%
  \BibitemOpen
  \bibfield  {author} {\bibinfo {author} {\bibfnamefont {M.}~\bibnamefont
  {Zilhão}}, \bibinfo {author} {\bibfnamefont {H.}~\bibnamefont {Witek}}, \
  and\ \bibinfo {author} {\bibfnamefont {V.}~\bibnamefont {Cardoso}},\ }\href
  {\doibase 10.1088/0264-9381/32/23/234003} {\bibfield  {journal} {\bibinfo
  {journal} {Class. Quant. Grav.}\ }\textbf {\bibinfo {volume} {32}},\ \bibinfo
  {pages} {234003} (\bibinfo {year} {2015})},\ \Eprint
  {http://arxiv.org/abs/1505.00797} {arXiv:1505.00797 [gr-qc]} \BibitemShut
  {NoStop}%
\bibitem [{\citenamefont {East}(2017)}]{East:2017mrj}%
  \BibitemOpen
  \bibfield  {author} {\bibinfo {author} {\bibfnamefont {W.~E.}\ \bibnamefont
  {East}},\ }\href {\doibase 10.1103/PhysRevD.96.024004} {\bibfield  {journal}
  {\bibinfo  {journal} {Phys. Rev.}\ }\textbf {\bibinfo {volume} {D96}},\
  \bibinfo {pages} {024004} (\bibinfo {year} {2017})},\ \Eprint
  {http://arxiv.org/abs/1705.01544} {arXiv:1705.01544 [gr-qc]} \BibitemShut
  {NoStop}%
\bibitem [{\citenamefont {Helfer}\ \emph {et~al.}(2019)\citenamefont {Helfer},
  \citenamefont {Aurrekoetxea},\ and\ \citenamefont {Lim}}]{Helfer:2018qgv}%
  \BibitemOpen
  \bibfield  {author} {\bibinfo {author} {\bibfnamefont {T.}~\bibnamefont
  {Helfer}}, \bibinfo {author} {\bibfnamefont {J.~C.}\ \bibnamefont
  {Aurrekoetxea}}, \ and\ \bibinfo {author} {\bibfnamefont {E.~A.}\
  \bibnamefont {Lim}},\ }\href {\doibase 10.1103/PhysRevD.99.104028} {\bibfield
   {journal} {\bibinfo  {journal} {Phys. Rev. D}\ }\textbf {\bibinfo {volume}
  {99}},\ \bibinfo {pages} {104028} (\bibinfo {year} {2019})},\ \Eprint
  {http://arxiv.org/abs/1808.06678} {arXiv:1808.06678 [gr-qc]} \BibitemShut
  {NoStop}%
\bibitem [{\citenamefont {Siemonsen}\ and\ \citenamefont
  {East}(2021)}]{Siemonsen:2020hcg}%
  \BibitemOpen
  \bibfield  {author} {\bibinfo {author} {\bibfnamefont {N.}~\bibnamefont
  {Siemonsen}}\ and\ \bibinfo {author} {\bibfnamefont {W.~E.}\ \bibnamefont
  {East}},\ }\href {\doibase 10.1103/PhysRevD.103.044022} {\bibfield  {journal}
  {\bibinfo  {journal} {Phys. Rev. D}\ }\textbf {\bibinfo {volume} {103}},\
  \bibinfo {pages} {044022} (\bibinfo {year} {2021})},\ \Eprint
  {http://arxiv.org/abs/2011.08247} {arXiv:2011.08247 [gr-qc]} \BibitemShut
  {NoStop}%
\bibitem [{\citenamefont {Clough}\ \emph {et~al.}(2022)\citenamefont {Clough},
  \citenamefont {Helfer}, \citenamefont {Witek},\ and\ \citenamefont
  {Berti}}]{Clough:2022ygm}%
  \BibitemOpen
  \bibfield  {author} {\bibinfo {author} {\bibfnamefont {K.}~\bibnamefont
  {Clough}}, \bibinfo {author} {\bibfnamefont {T.}~\bibnamefont {Helfer}},
  \bibinfo {author} {\bibfnamefont {H.}~\bibnamefont {Witek}}, \ and\ \bibinfo
  {author} {\bibfnamefont {E.}~\bibnamefont {Berti}},\ }\href@noop {} {\
  (\bibinfo {year} {2022})},\ \Eprint {http://arxiv.org/abs/2204.10868}
  {arXiv:2204.10868 [gr-qc]} \BibitemShut {NoStop}%
\bibitem [{\citenamefont {{Dedner}}\ \emph {et~al.}(2002)\citenamefont
  {{Dedner}}, \citenamefont {{Kemm}}, \citenamefont {{Kr{\"o}ner}},
  \citenamefont {{Munz}}, \citenamefont {{Schnitzer}},\ and\ \citenamefont
  {{Wesenberg}}}]{2002JCoPh.175..645D}%
  \BibitemOpen
  \bibfield  {author} {\bibinfo {author} {\bibfnamefont {A.}~\bibnamefont
  {{Dedner}}}, \bibinfo {author} {\bibfnamefont {F.}~\bibnamefont {{Kemm}}},
  \bibinfo {author} {\bibfnamefont {D.}~\bibnamefont {{Kr{\"o}ner}}}, \bibinfo
  {author} {\bibfnamefont {C.~D.}\ \bibnamefont {{Munz}}}, \bibinfo {author}
  {\bibfnamefont {T.}~\bibnamefont {{Schnitzer}}}, \ and\ \bibinfo {author}
  {\bibfnamefont {M.}~\bibnamefont {{Wesenberg}}},\ }\href {\doibase
  10.1006/jcph.2001.6961} {\bibfield  {journal} {\bibinfo  {journal} {Journal
  of Computational Physics}\ }\textbf {\bibinfo {volume} {175}},\ \bibinfo
  {pages} {645} (\bibinfo {year} {2002})}\BibitemShut {NoStop}%
\bibitem [{\citenamefont {Pretorius}(2005)}]{Pretorius:2004jg}%
  \BibitemOpen
  \bibfield  {author} {\bibinfo {author} {\bibfnamefont {F.}~\bibnamefont
  {Pretorius}},\ }\href {\doibase 10.1088/0264-9381/22/2/014} {\bibfield
  {journal} {\bibinfo  {journal} {Class. Quant. Grav.}\ }\textbf {\bibinfo
  {volume} {22}},\ \bibinfo {pages} {425} (\bibinfo {year} {2005})},\ \Eprint
  {http://arxiv.org/abs/gr-qc/0407110} {arXiv:gr-qc/0407110 [gr-qc]}
  \BibitemShut {NoStop}%
\end{thebibliography}%

\appendix
\section{Evolution Equations} 
We evolve the coupled Abelian Higgs equations for the vector and complex scalar field (Eq.~\ref{eqn:eom}) on a
black hole background. These equations have a $U(1)$ symmetry and are invariant under
the gauge transformation
\beqa
\Phi \rightarrow \Phi e^{-i \theta} , \
A^a \rightarrow A^a-\frac{1}{g} \nabla^a \theta . 
\eeqa
In unitary gauge, $\Phi$ is chosen to be real. However, this gauge can be problematic when vortices
form, and we instead evolve the equations 
using the Lorenz gauge $\nabla_a A^a=0$.
 
For the evolution of the scalar field, we directly evolve the real and imaginary components of $\Phi=\Phi_R+i\Phi_I$
 and $\partial_t \Phi$ according to
\beq
\Box \Phi = ig\Phi \nabla_a A^a +  2ig A^a \nabla_a \Phi +  g^2 A_aA^a\Phi + \lambda(|\Phi|^2-v^2)\Phi \ . 
\label{eqn:scalar}
\eeq
Choosing the Lorenz gauge, the first term on the right hand side vanishes. 

As in Refs.~\cite{Zilhao:2015tya,East:2017mrj,Helfer:2018qgv}, to evolve
the vector field we decompose into 
time and spatial components:
\beq
A_a = \chi_a+n_a\chi,
\eeq
where $n_a$ is the unit normal to slices of constant coordinate
time, and introduce an electric field
\beq
E_i = \gamma^{a}_i F_{ab} n^b,
\eeq
where $\gamma^a_b=\delta^a_b+n^an_b$ is the spatial projection operator.  
Here, and in the following, the indices $\{i,\ j,\ k,\ \ldots\}$ are spatial
indices that run from 1 to 3, as opposed to the spacetime indices $\{a,\ b,\ c,\ \ldots\}$,
which run from 0 to 3.
Following Refs.~\cite{2002JCoPh.175..645D,Zilhao:2015tya}, we also introduce an auxiliary field $Z$ designed to damp away violations
of the constraint.
In terms of these variables, the evolution equations are
\beqa
N^{-1}(\partial_t - \mathcal{L}_{\beta}) \chi_i &=& -E_i-\partial_i \chi -\chi \partial_i \log N,\\
N^{-1}(\partial_t - \mathcal{L}_{\beta}) \chi &=& K\chi -\mathcal{D}_i \chi^i -\chi^i \partial_i \log N -Z,\\
N^{-1}(\partial_t - \mathcal{L}_{\beta}) E^i &=& K E^i +\mathcal{D}^iZ+\epsilon^{ijk}\mathcal{D}_jB_k\nonumber \\ && -\epsilon^{ijk}B_j\partial_k \log N +g^2|\Phi|^2 \chi^i \nonumber \\ 
&&-g\left( \Phi_R \partial^i \Phi_I-\Phi_I \partial^i \Phi_R \right),\\
N^{-1}(\partial_t - \mathcal{L}_{\beta}) Z &=& 
-\sigma Z + \mathcal{D}_i E^i +g^2|\Phi|^2 \chi \nonumber \\
 && 
+gN^{-1} 
\Phi_R(\partial_t-\beta^i \partial_i)\Phi_I \nonumber \\
&&-gN^{-1}\Phi_I(\partial_t-\beta^i \partial_i)\Phi_R ,
\label{eqn:vec}
\eeqa
where $N$ and $\beta^i$ are the lapse and shift, respectively, $K$ is the trace of the
extrinsic curvature, $\mathcal{D}_i$ is the covariant derivative associated with the
spatial metric, $\epsilon_{ijk}$ is the spatial totally antisymmetric
tensor, and $B^i=\epsilon^{ijk}\mathcal{D}_j \chi_k$ is the magnetic field. The coupled vector-scalar fields are evolved on a black hole spacetime in Kerr-Schild coordinates~\cite{1965cngg.conf..222K}. 
In the numerical evolution scheme, spatial derivatives are calculated with
standard fourth-order finite difference stencils and the time evolution is
carried out with fourth-order Runge-Kutta, though the interpolation in time for
mesh refinement boundaries is only third-order
accurate~\cite{East:2017mrj,Siemonsen:2020hcg}. 

\section{Initial Conditions}
We construct initial data for our evolutions by first evolving an approximate
symmetry-reduced version of our model. In this approximate version, we assume
that the vector field has an $m=1$ azimuthal symmetry, and that the scalar field
is real (i.e. we take unitary gauge) and has an $m=0$ azimuthal symmetry.
This means the Lorenz gauge condition is replaced by $\nabla_a A^a=-2A^a\nabla_a \log \Phi$. 
In order for this to be consistent, we also have to modify the scalar equation of
motion Eq.~\ref{eqn:scalar} by replacing the $g^2 A^2 \Phi$ term by its azimuthally-averaged value.
Similar to Ref.~\cite{East:2017mrj}, we choose an initial vector perturbation
and then evolve the symmetry-reduced system for a number of e-folding times
until the solution is dominated by the fastest growing superradiantly unstable
mode.  The result is then taken as initial conditions for evolving the full
system without symmetry assumptions.  There is a short initial transient due to
the fact that the scalar field must relax to its non-axisymmetric and Lorenz
gauge value. However, as can be seen from, e.g., Fig.~\ref{fig:field_values}, this is mild as we consider initial 
conditions with  $\min(|\Phi|/v)>0.9$.

\section{Numerical Convergence} 
For our computational domain, we use six cubic levels of mesh refinement centered on the black hole.
The finest level has a linear dimension of $L\approx 5 M$, and each subsequent coarser level
has a linear dimension and grid spacing that is twice as large. 
We use compactified coordinates which extend to spatial infinity following Ref.~\cite{Pretorius:2004jg}.
For our default resolution, the finest level has a grid spacing of $dx\approx 0.1 \lambda^{-1/2}v^{-1}$. 

\begin{figure}
\begin{center}
\includegraphics[width=\columnwidth,draft=false]{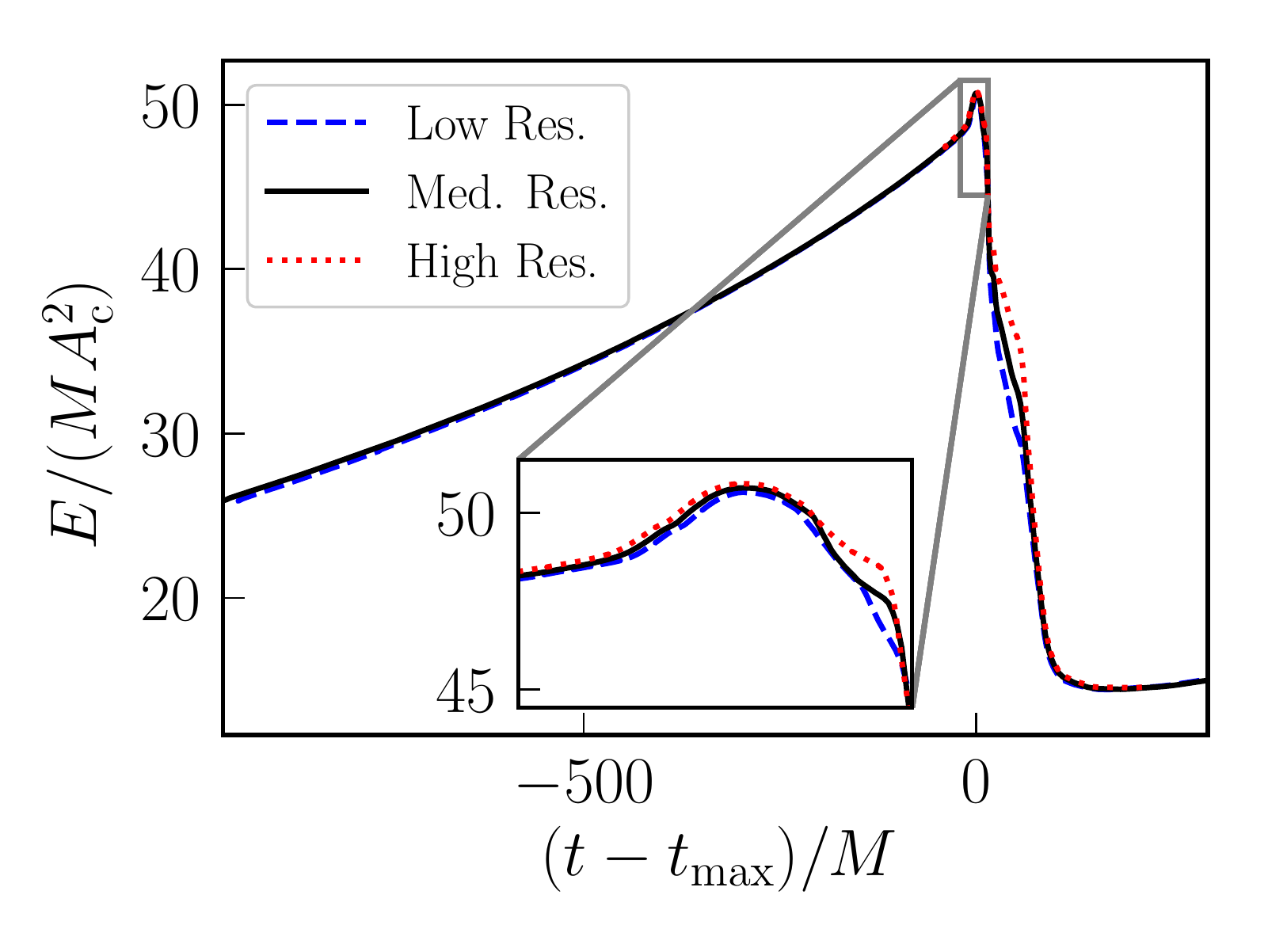}
\end{center}
\caption{
    Energy as a function of time for $\alpha=0.4$, $\lambda/g^2=25$, and three numerical resolutions.
    The inset shows a zoom-in around the time the energy reaches a maximum.
\label{fig:ej_conv}
}
\end{figure}

For the case with $\alpha=0.4$ and $\lambda/g^2=25$, we perform a resolution
study with grid spacing
that is $0.75\times$ and $0.5\times$ as large.  For computational expediency,
for the highest resolution we begin $\sim 50 M$ before the system achieves peak
energy, using the next highest resolution to set initial conditions.  We show
the total energy as a function of time for the three resolutions in
Fig.~\ref{fig:ej_conv}.
Using the three resolutions, we estimate that the lowest resolution (which is
the minimum resolution for all the results in the main text) underestimates the
maximum energy by $\approx 1\%$.  (The Richardson extrapolation of this
diagnostic quantity is consistent with first order convergence, likely due to
the discrete way points inside the black hole horizon are excluded when
numerically integrating the total energy.)

\end{document}